\documentclass[fleqn,usenatbib]{mnras}

\usepackage{newtxtext,newtxmath}
\usepackage[T1]{fontenc}

\DeclareRobustCommand{\VAN}[3]{#2}
\let\VANthebibliography\thebibliography
\def\thebibliography{\DeclareRobustCommand{\VAN}[3]{##3}\VANthebibliography}

\usepackage{graphicx}	% Including figure files
\usepackage{amsmath}	% Advanced maths commands
\usepackage{chemformula}
\usepackage[utf8]{inputenc}
\usepackage{url}
\newcommand{\taubootisb}{$\tau$ Bo\"{o}tis b}

\newcommand{\KpVsys}{K$_\mathrm{P}$ - V$_\mathrm{sys}$}

\newcommand{\Kp}{K$_\mathrm{P}$}
\newcommand{\Vsys}{V$_\mathrm{sys}$}

\newcommand{\HtwoO}{$\mathrm{H_2O}$}
\newcommand{\HtwoS}{$\mathrm{H_2S}$}
\newcommand{\CO}{$\mathrm{CO}$}

\newcommand{\COtwo}{$\mathrm{CO_2}$}
\newcommand{\OH}{$\mathrm{OH}$}

\newcommand{\logWater}{$\log_{10}(\mathrm{H_2O})$}

\newcommand{\logZpByZs}{log$_{10}$[Z$_{\mathrm{P}}$/Z$_{\odot}$]}

\newcommand{\kms}{\ensuremath{\mathrm{km\,s^{-1}}}}

\newcommand{\FpFs}{F$_\mathrm{P}$/F$_\mathrm{S}$}
\newcommand{\Fp}{F$_\mathrm{P}$}
\newcommand{\Fs}{F$_\mathrm{S}$}
\newcommand{\PT}{\ensuremath{\mathrm{P-T}}}

\newcommand{\Npca}{\ensuremath{\mathrm{N_{PCA}}}}

\title[Metal depleted atmosphere of WASP-122b]{The Roasting Marshmallows Program with IGRINS on Gemini South III: Seeing deeper into the metal depleted atmosphere of a gas-giant on the cusp of the hot to ultra-hot Jupiter transition}

\author[Panwar et al.]{
Vatsal Panwar$^{1,2}$\thanks{E-mail: vatsal.panwar@warwick.ac.uk},
Matteo Brogi$^{3,4}$,
Krishna Kanumalla$^{5}$,
Michael R. Line$^{5}$,
Siddharth Gandhi$^{1,2}$,
\newauthor
Peter C.B. Smith$^{5}$,
Jacob L. Bean$^{6}$,
Lorenzo Pino$^{9}$,
Arjun B. Savel$^{7}$,
Joost P. Wardenier$^{8}$,
Heather Cegla$^{1,2}$\thanks{UKRI Future Leaders Fellow},
\newauthor
Hayley Beltz$^{7}$,
Megan Weiner Mansfield$^{5,7}$,
Jorge A. Sanchez$^{5}$,
Jean-Michel Désert$^{10}$,
Luis Welbanks$^{5}$,
\newauthor
Viven Parmentier$^{11}$,
Changwoo Kye$^{12}$, 
Jonathan J. Fortney$^{13}$, 
Tomás de Azevedo Silva$^{9}$
\\
$^{1}$Department of Physics, University of Warwick, Coventry, UK, CV47AL\\
$^{2}$Center for Exoplanets and Habitability, University of Warwick, Coventry, UK, CV47AL\\
$^{3}$Dipartimento di Fisica, Università degli Studi di Torino, via P. Giuria 1, Turin, I-10125, Italy \\
$^{4}$Osservatorio Astrofisico di Torino, INAF, via Osservatorio 20, Pino Torinese, I-10025, Italy \\
$^{5}$School of Earth \& Space Exploration, Arizona State University, Tempe AZ 85287, USA \\
$^{6}$Department of Astronomy \& Astrophysics, University of Chicago, Chicago, IL, USA \\
$^{7}$Department of Astronomy, University of Maryland, College Park, 4296 Stadium Dr., College Park, MD 207842 USA \\
$^{8}$Institut Trottier de Recherche sur les Exoplanètes, Université de Montréal, Montréal, Québec, H3T 1J4, Canada\\
$^{9}$INAF - Osservatorio Astrofisico di Arcetri, Florence, Italy\\
$^{10}$Anton Pannekoek Institute for Astronomy, University of Amsterdam, P.O. Box 94249, Noord Holland, NL-1090GE Amsterdam, the Netherlands\\
$^{11}$Laboratoire Lagrange, Observatoire de la Côte d’Azur, CNRS, Université Cóte d’Azur, Nice, France \\
$^{12}$Department of Physics and Astronomy, Seoul National University, Seoul 08826, Korea \\
$^{13}$Department of Astronomy and Astrophysics University of California, Santa Cruz, CA, USA
}

\date{Accepted 2025 July 07. Received 2025 July 02; in original form 2025 March 19}

\pubyear{\the\year{}}

\begin{document}
\label{firstpage}
\pagerange{\pageref{firstpage}--\pageref{lastpage}}
\maketitle

\begin{abstract}
Ultra-hot Jupiters (dayside temperatures T$_{\mathrm{day}}$>2200 K) are a class of gas-giant exoplanets that, due to extreme stellar irradiation, show a peculiar combination of thermochemical properties in the form of molecular dissociation, atomic ionization, and inverted thermal structures. Atmospheric characterization of gas giants lying in the transitional regime between hot and ultra-hot Jupiters can help in understanding the physical mechanisms that cause the fundamental thermochemical transition in atmospheres between the two classes of hot gas giants. Using high-resolution cross-correlation spectroscopy with the IGRINS spectrograph on Gemini South (1.4 to 2.5 $\mu$m), we present the day-side high-resolution spectrum of WASP-122b (T$_{\mathrm{day}}$=2258$ \pm$ 54 K), a gas-giant situated at this transition. We detect the signal from \HtwoO{}, based on which we find that WASP-122b has a significantly metal-depleted atmosphere with metallicity log$_{10}$[Z$_{\mathrm{P}}$/Z$_{\odot}$] = $-$1.48±0.25 dex (0.033$_{-0.016}^{+0.018}$ $\times$ solar), and solar/sub-solar C/O ratio = 0.36±0.22 (3$\sigma$ upper limit 0.82). Drastically low atmospheric metallicity pushes the contribution function to higher pressures, resulting in the planetary spectral lines to originate from a narrow region around 1 bar where the thermal profile is non-inverted. This is inconsistent with solar composition radiative convective equilibrium (RCTE) which predicts an inverted atmosphere with spectral lines in emission. The measured sub-solar metallicity and solar/sub-solar C/O ratio is inconsistent with expectations from core-accretion. We find the planetary signal to be significantly shifted in K$_{\mathrm{P}}$ and V$_{\mathrm{sys}}$, which is in tension with the predictions from global circulation models and require further investigation. Our results highlight the detailed information content of high-resolution spectroscopy data and their ability to constrain complex atmospheric thermal structures and compositions of exoplanets.
% , 
% Our results highlight the detailed information content of ground-based high-resolution spectroscopy data, and their ability to constrain complex thermal structures of gas-giant exoplanets. We discuss the implications of our findings on the understanding of the physical mechanisms at play at the cusp of transition from hot to ultra-hot Jupiters, and for the formation history of WASP-122b in particular.
 
\end{abstract}


%%%%%%%%%%%%%%%%%%%%%%%%%%%%%%%%%%%%%%%%%%%%%%%%%%

%%%%%%%%%%%%%%%%% BODY OF PAPER %%%%%%%%%%%%%%%%%%

\section{Introduction}
The atmospheric thermal structures of irradiated gas-giant exoplanet atmospheres are determined by the interplay of stellar irradiation, interior heat flux, and the opacity of atomic and molecular species in the atmosphere, which together govern the energy balance of the atmosphere \citep{seager_extrasolar_1998, barman_phase-dependent_2005, sudarsky_theoretical_2003, burrows_theoretical_2007}.  The atmosphere of a hot gas-giant heats up by absorbing the incident stellar radiation typically in the visible wavelengths due to alkali (Na, K) and other refractory bearing species (e.g. TiO, VO), and cools by emitting radiation in the infrared wavelengths from molecules like \HtwoO{}, CO, and \COtwo{}. The resulting variation of temperature with altitude or pressure, often referred to as the pressure temperature (\PT{}) profile, is intimately connected to the atmospheric metallicity which controls the opacity of the atmosphere \citep{marley_cool_2015}. Studies by \cite{hubeny_possible_2003, fortney_atmosphere_2006, fortney_unified_2008} and more recently 3D global circulation model (GCM) simulations for a population of hot-Jupiters by \cite{roth_hot_2024} have indicated that a fundamental change in the atmospheric opacity happens in the planetary equilibrium temperature range of $\sim$ 1700 - 2000 K when TiO and VO, the main visible stellar light absorbers, exist in gas-phase and heat up the upper atmosphere causing thermal inversion (increasing temperature with altitude). 

Ultra-hot Jupiters \citep{arcangeli_h-_2018, mansfield_hstwfc3_2018, kreidberg_global_2018} are a class of even hotter gas-giants (T$_{\mathrm{day}}$ $>$ 2200 K) for which a combination of high stellar irradiation and low-surface gravity causes most molecules, including TiO, VO and \HtwoO{}, to be thermally dissociated \citep{parmentier_thermal_2018}. The role of heating up the atmosphere in case of ultra-hot Jupiters is expected to be taken over by atomic species, e.g., resonant lines of Na I, K I, Mg I, Ca I at deeper pressures, and by Fe I and Fe II at lower pressures \citep{parmentier_thermal_2018}. The prevalence of thermally inverted \PT{} profiles for ultra-hot Jupiters has been confirmed from numerous measurements of their emission spectra \citep{arcangeli_h-_2018, mansfield_hstwfc3_2018, pino_neutral_2020, mikal-evans_diurnal_2022, prinoth_titanium_2022, van_sluijs_carbon_2023, smith_roasting_2024, pelletier_crires_2024, deibert_high-resolution_2024}.

A combination of factors including surface gravity, equilibrium temperature, atmospheric composition, and stellar spectral type are expected to affect the onset and strength of thermal inversions across the hot and ultra-hot Jupiter population. For example, increasing the host star temperature \citep{lothringer_influence_2019} and atmospheric metallicity \citep{parmentier_thermal_2018} can both cause steeper (higher $\Delta$T/$\Delta$P gradient of the temperature (T) - pressure (P) profile) thermal inversions. Analysis of a statistical sample of secondary eclipses of hot and ultra-hot Jupiters obtained through Spitzer Infared Array Camera (IRAC) in the 3.6 and 4.5 $\mu$m photometric bandpasses and using the strength of CO emission feature as a proxy indicated that the transition to thermally inverted \PT{} profile happens around 1700 K \cite{baxter_transition_2020, garhart_statistical_2020}. Investigations into the emission spectra of a sample of 19 hot and ultra-hot Jupiters obtained using Wide Field Camera 3 on the Hubble Space Telescope (HST/WFC3) by \cite{mansfield_unique_2021} revealed a scatter in the strength of the 1.4 $\mu$m \HtwoO{} feature around the prediction from self-consistent atmospheric models, which could be explained as an outcome of rain-out of TiO/VO (for cooler hot-Jupiters) and variations in atmospheric metallicity and C/O ratio (for hotter hot-Jupiters).  

\begin{figure}
\centering
\includegraphics[width=0.5\textwidth]{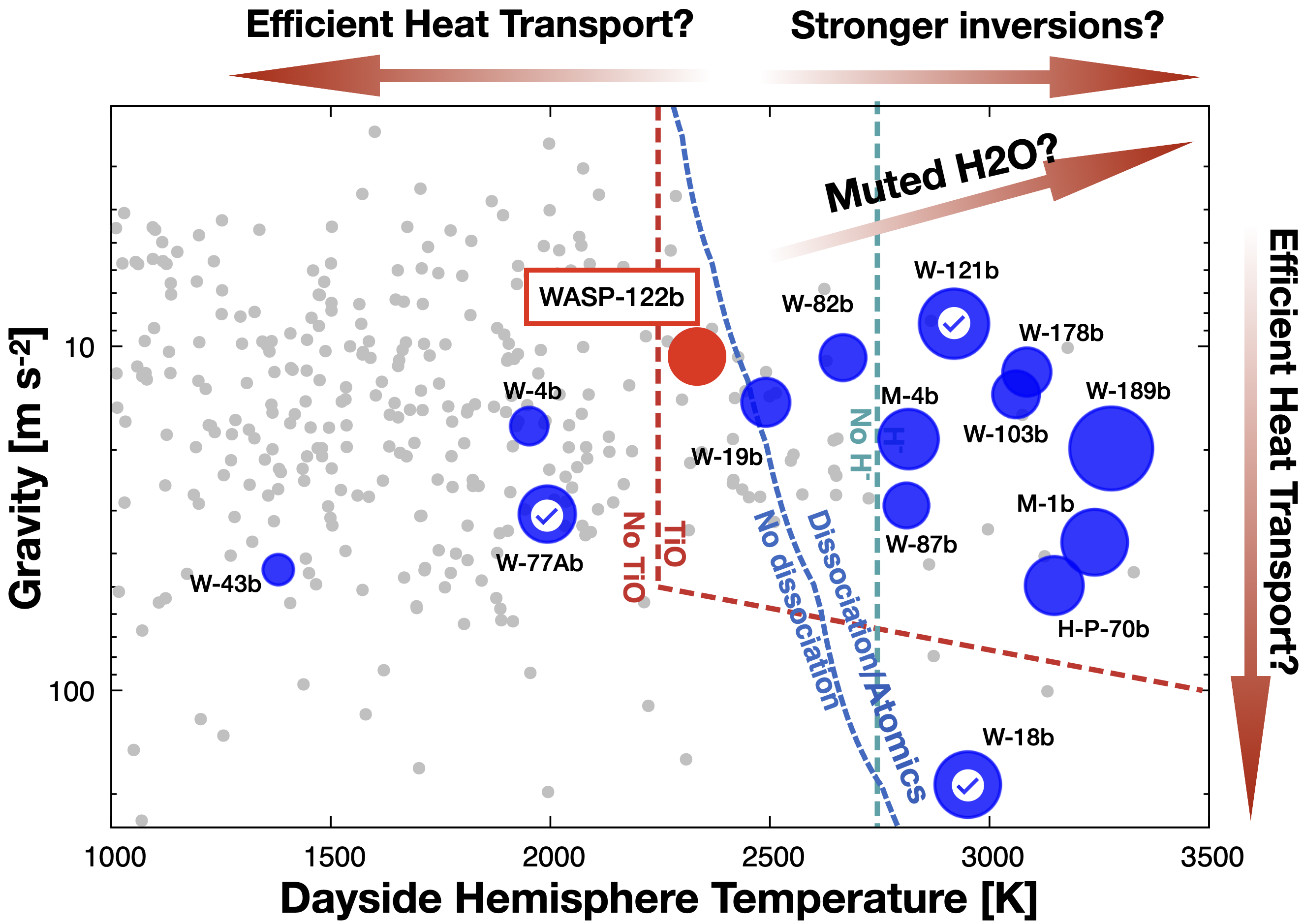}
\label{fig:roasting_marshmallows}
\caption{
Surface gravity vs dayside temperature parameter space explored by the Roasting Marshmallows program to measure the dayside emission spectra of hot to ultra-hot Jupiters with IGRINS on Gemini South. General planet population is marked by grey symbols, and the planet names observed by the program in blue (W = WASP, and M = MASCARA, H-P = HAT-P). The symbol size is proportional to the S/N expected for each target relative to WASP-77 Ab. All the planets targeted by this survey span atmospheric transitions predicted by \protect\cite{parmentier_thermal_2018} including onset of thermal inversions by TiO, and molecular thermal dissociation. The dayside temperatures plotted here were calculated assuming redistribution efficiency {\it f} predicted from the trend measured by \protect\cite{parmentier_cloudy_2021}. Observations for planets marked by a white tick have been published already (\protect\cite{line_solar_2021, brogi_roasting_2023, smith_roasting_2024}). The focus of this work is the hot Jupiter WASP-122b (T$_{\mathrm{day}}$ = 2258 $\pm$ 54 K for {\it f} = 1.98) which is situated in the transition regime between hot and ultra-Jupiter where the onset of thermal inversion is expected due to TiO/VO and molecular dissociation.
}
\end{figure}

There are ample theoretical predictions and corresponding observational evidence from broadband photometry and low to high resolution spectroscopy on the prevalence and causes for thermal inversion in the atmospheres of ultra-hot Jupiters. However, there are few observational constraints on the thermal structures of gas-giants in the parameter space between hot to ultra-hot Jupiters where this fundamental transition likely happens. WASP-122b/KELT-14b \citep{rodriguez_kelt-14b_2016, turner_wasp-120_2016} (hereafter referred to as WASP-122b) is one such hot-Jupiter (T$_{\mathrm{eq}}$ $\sim$ 1904 $\pm$ 54 K, T$_{\mathrm{day}}$ = 2258 $\pm$ 54 K for redistribution efficiency \textit{f} = 1.98) situated in this transition regime in the day-side temperature vs surface gravity space (Figure \ref{fig:roasting_marshmallows}). Moreover, given its short orbital period (1.71 day), inflated size (1.52 R$_{\mathrm{Jup}}$, 1.19 M$_{\mathrm{Jup}}$), and bright (K mag = 9.4) host star (T$_{\mathrm{eff}}$ = 5800 K, spectral type G4), it is an ideal candidate for emission spectroscopy (see Table \ref{tab:system_params} for system properties). It is a particularly favourable target for constraining its atmospheric thermal structure and metallicity through high-resolution cross-correlation spectroscopy (HRCCS; \cite{snellen_orbital_2010, brogi_signature_2012, birkby_detection_2013}). HRCCS uses high-resolution (R$>$15000) time-series spectra and leverages the Doppler shift of many spectral lines induced by its orbital motion to disentangle the planetary atmospheric signal from telluric and stellar contamination. The technique is sensitive to line-shapes and contrasts (\cite{birkby_exoplanet_2018}) and has been proven to obtain statistical constraints \citep{brogi_retrieving_2019, gibson_detection_2020} on the vertical thermal structure and atmospheric compositions of hot to ultra-hot Jupiters (e.g. WASP-77 Ab \cite{line_solar_2021}; WASP-76b \cite{gandhi_spatially_2022, weiner_mansfield_metallicity_2024, pelletier_vanadium_2023}; \taubootisb{} \cite{pelletier_where_2021, panwar_mystery_2024}; WASP-121b \citep{smith_roasting_2024, pelletier_crires_2025}). The precise constraints on elemental ratios of refractory and volatile species enabled by HRCCS have enabled linking the compositions of planetary atmospheres with their formation histories \citep{line_solar_2021, smith_roasting_2024, pelletier_crires_2025}. Phase resolved analyses of HRCCS signals have also revealed expected 3D signatures of dynamical processes like atmospheric circulation and drag in ultra-hot Jupiters (\cite{beltz_significant_2021, wardenier_phase-resolving_2024}). 

In this work, we present the emission spectroscopy of the transitional hot Jupiter WASP-122b in the infrared H and K bands (1.4 to 2.4 $\mu$m) using the Immersion GRating INfrared Spectrometer (IGRINS, \cite{park_design_2014, mace_igrins_2018} on Gemini-South) with the goal to constrain its atmospheric composition and thermal structure. We use our constraints on WASP-122b to investigate the role of atmospheric composition and stellar irradiation on the onset of thermal inversions in the regime between hot and ultra-hot Jupiters. We describe our observations and data reduction and analysis steps in Section \ref{sec:observations}, followed by steps used to search for planetary spectral signatures and constraining the atmospheric abundances and \PT{} profile in Section \ref{sec:cross_correlation_analysis} and \ref{sec:retrieval_analysis} respectively. We discuss our findings and their implications in the context of thermal inversions in hot to ultra-hot Jupiters, and for the potential formation history of WASP-122b in Section \ref{sec:discussion}, and summarize our conclusions in Section \ref{sec:conclusions}.

\section{Spectral Extraction and Telluric and Stellar Detrending}
\label{sec:observations}

%% System properties 
\begin{table*}
    \centering
    \renewcommand{\arraystretch}{1.2}
    \caption{Stellar and planetary parameters for WASP-122b/KELT-14b from \protect\cite{rodriguez_kelt-14b_2016} and \protect\cite{turner_wasp-120_2016}. The parameters from both are consistent with each other except the stellar radius and mass and hence the planetary radius and mass. This likely arises from the differences in stellar spectral analysis between the two studies. We use values from \protect\cite{rodriguez_kelt-14b_2016} for this work, and find that using values from \protect\cite{turner_wasp-120_2016} do not change our results significantly.}
    \begin{tabular}{c c c}
     \hline
      \hline
      Stellar Parameters  &  &  \\
     \hline
     Name & \cite{rodriguez_kelt-14b_2016} & \cite{turner_wasp-120_2016}  \\
     \hline
        R$_\mathrm{star}$ & 1.368 $\pm$ 0.07 [R$_\odot$] & 1.52 $\pm$ 0.03 [R$_\odot$]  \\
        M$_\mathrm{star}$ & 1.178 $\pm$ 0.06 [M$_\odot$] & 1.239 $\pm$ 0.04 [M$_\odot$]  \\
        log g$_\mathrm{star}$ & 4.23 $\pm$ 0.04 [log$_{10}$[cgs]] & 4.166 $\pm$ 0.016 [log$_{10}$[cgs]] \\
        T$_\mathrm{eff}$ & 5802 $\pm$ 95 [K] & 5750 $\pm$ 120 [K] \\
        K mag & 9.424 $\pm$ 0.023 & -- \\
        V$_\mathrm{sys}$ & 34.62 $\pm$ 0.13 [km s$^{-1}$] & 34.59 $\pm$ 0.002 [km s$^{-1}$]  \\
        $v_\mathrm{S}$ sin$i$ & 7.7 $\pm$ 0.4 [km s$^{-1}$] & 3.3 $\pm$ 0.8 [km s$^{-1}$]   \\
        $\mathrm{[Fe/H]}$ & 0.326 $\pm$ 0.09 & 0.32 $\pm$ 0.09 \\
        % $\mathrm{[O/H]}$ & 0.42 $\pm$ 0.07 &  \\
        % $\mathrm{[C/H]}$ & 0.04 $\pm$ 0.05 &  \\
        % $\mathrm{[Fe/H]}$ & 0.24 $\pm$ 0.03 &  \\
        % $\mathrm{[Mg/H]}$ & 0.15 $\pm$ 0.04 &  \\
        % $\mathrm{[Ca/H]}$ & 0.25 $\pm$ 0.03 & \\
        % $\mathrm{[Si/H]}$ & 0.24 $\pm$ 0.03 & \\
        % $\mathrm{[V/H]}$ & 0.01 $\pm$ 0.06 & \\
        % $\mathrm{[Ti/H]}$ & 0.22 $\pm$ 0.04& \\
        % $\mathrm{[Cr/H]}$ & 0.23 $\pm$ 0.04 & \\
        % C/O & 0.23 $\pm$ 0.05 & '' \\
     \hline
      \hline
      Planet Parameters (Literature)  & &  \\
     \hline
        R$_\mathrm{P}$ &  1.52 $\pm$ 0.11 [R$_\mathrm{Jup}$] & 1.74 $\pm$ 0.05 [R$_\mathrm{Jup}$] \\
        M$_\mathrm{P}$ &  1.196$\pm$ 0.07 [M$_\mathrm{Jup}$] & 1.284$\pm$ 0.03 [M$_\mathrm{Jup}$] \\
        log g$_\mathrm{P}$ &  3.107$\pm$ 0.07 [log$_{10}$[cgs]] & 3.02$\pm$ 0.01 [log$_{10}$[cgs]] \\        
        T$_\mathrm{eq}$ & 1904 $\pm$ 54 [K] & 1970 $\pm$ 50 [K]  \\
        T$_0$ & 2457091.028632 $\pm$ 4 $\times$ 10$^{-4}$ [BJD$_\mathrm{TDB}$] & 2456665.22401 $\pm$ 2.1 $\times$ 10$^{-4}$ [BJD$_\mathrm{TDB}$] \\
        P$_\mathrm{orb}$ & 1.7100588 $\pm$ 2.5$\times 10^{-6}$ [day] & 1.7100566 $\pm$ 3.2$\times 10^{-6}$ [day] \\
        a & 0.02956 $\pm$ 0.0005 [A.U.] & 0.03005 $\pm$ 0.0003 [A.U.]  \\
        $i$ & 79.67 $\pm$ 0.8 [$^{\circ}$]  & 78.3 $\pm$ 0.3 [$^{\circ}$] \\
        (R$_\mathrm{P}$/R$_\mathrm{S}$)$^2$ & 0.01306 $\pm$ 0.0006 & 0.01386 $\pm$ 0.00029 \\
     \hline
      Planet Parameters (Derived)  & &  \\
     \hline
        K$_\mathrm{P}$ & 188.05 $\pm$ 3.2 [km s$^{-1}$] & 191.17 $\pm$ 1.9 [km s$^{-1}$]   \\
        $v_\mathrm{P}$ sin$i$ & 4.54 $\pm$ 0.33 [km s$^{-1}$] & 5.19 $\pm$ 0.15 [km s$^{-1}$]  \\
        \hline 
        \hline
        
    \end{tabular}
    \label{tab:system_params}
\end{table*}

We observed WASP-122b for a single night on UTC 4th December 2023 with IGRINS on Gemini South as part of the Large-and-Long Program “Roasting Marshmallows: Disentangling Composition and Climate in Hot Jupiter Atmospheres through High-Resolution Thermal Emission Cross-Correlation Spectroscopy” (GS-2023B-LP-206, PI M. Line). The main goal of the program has been to conduct high spectral resolution phase- resolved cross- correlation spectroscopy of 15 hot Jupiters covering dayside temperatures 1400 - 2600 K, and diagnose the chemical, thermal, and dynamical mechanisms that drive key transitions in highly irradiated atmospheres. The observations of WASP-122b spanned 4.75 hours covering the pre-eclipse phase range 0.36 to 0.47 (corresponding to T$_0$ = 2458654.016813 $\pm$ 7 $\times$ 10$^{-5}$ BJD$_\mathrm{TDB}$ and orbital period P$_\mathrm{orb}$ = 1.71005328 $\pm$ 1.4 $\times$ 10$^{-7}$ from the most recent ephemeris measured by \cite{kokori_exoclock_2023}). The observations were obtained as a continuous sequence of 39 AB pairs exposures (in ABBA nodding pattern, 324 seconds per AB sequence, which includes the exposure time for each A and B position and the overheads). The median signal-to-noise ratio (SNR) obtained per AB exposure was 70 and 65 in H and K bands.

The 1D spectra from all the exposures were extracted from the raw data using the IGRINS Pipeline Package \citep{lee_plp_2016, mace_igrins_2018}. Following the same steps as \cite{line_solar_2021, brogi_roasting_2023, smith_roasting_2024}, we apply additional corrections to the reduced 1D spectrum for each exposure, which involved trimming 100 pixels at start and end of each order, substituting any negative or NaN values from bad pixels to zero, removing orders heavily saturated by telluric absorption ($<$1.45 $\mu$m, between 1.79 $\mu$m and 1.97 $\mu$m, and $>$ 2.42 $\mu$m). We re-aligned the spectrum for each exposure to the wavelength solution of the exposure with the highest SNR (hereafter reference spectrum). This was done by computing the least squares fit between each spectrum (with linear shift and stretch for the wavelength solution as free parameters for each exposure) and the reference spectrum. For each exposure, we applied the best fit linear shift and stretch coefficients to the reference wavelength solution, and re-aligned the spectrum to this using B-spline interpolation. This yielded spectral datacubes with dimensions N$_{\mathrm{order}}$ $\times$ N$_{\mathrm{frame}}$ $\times$ N$_{\mathrm{pixel}}$, where N$_{\mathrm{order}}$ = 42, N$_{\mathrm{frame}}$ = 39, and N$_{\mathrm{pixel}}$ = 1848. 

\begin{figure}
\centering
\includegraphics[width=0.5\textwidth]{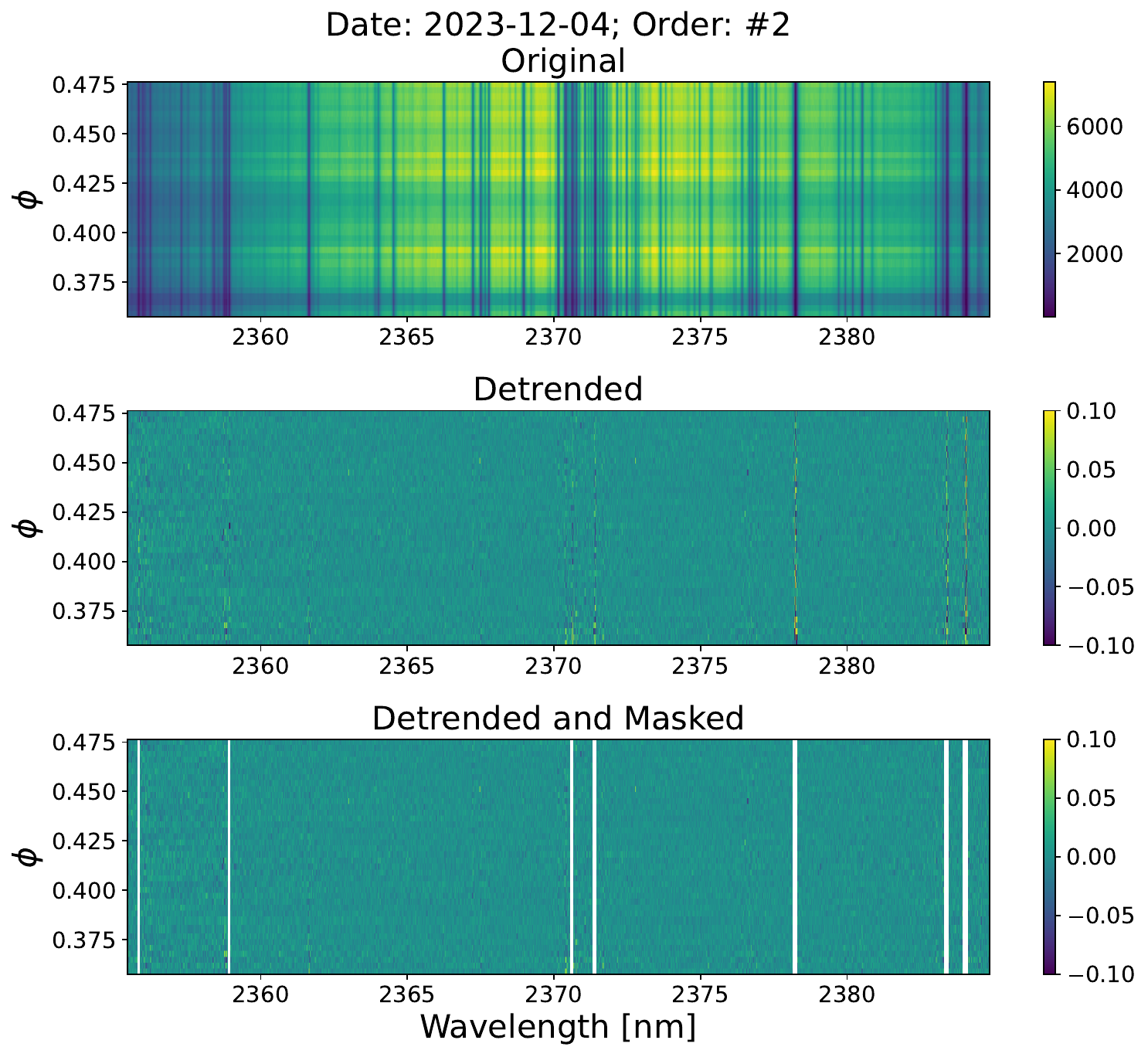}
\caption{Demonstration of detrending the data to remove the contamination from stellar and telluric lines before cross-correlation. The top panel shows the original raw flux data cube where each row is the 1D spectrum for an exposure, the horizontal axis is the wavelength, and the vertical axis is the orbital phase for each exposure. The middle panel is the PCA detrended data cube. The bottom panel shows the same detrended data cube in the middle panel, but the spectral channels that still retain statistically significant level of residual noise have been masked out in white along with the bad pixels. The color bar in each panel represents measured flux in arbitrary linear units. 
}
\label{fig:pca_detrending}
\end{figure}

We use principal component analysis (PCA) on an order by order basis to detrend the data and remove the telluric and stellar contamination. We follow the same procedure as described in \cite{panwar_mystery_2024}, which we describe here in brief. We first standardize the datacubes of each order (N$_{\mathrm{frame}}$ $\times$ N$_{\mathrm{pixel}}$) by subtracting the mean flux and normalizing by the standard deviation for each spectral column. This ensures that noisier telluric channels are not overly weighted as compared to the rest of the spectral channels, while still letting PCA capture the required information to model the telluric channels. We then apply a Singular Value Decomposition (SVD) to the standardized datacube and compute a set of eigenvectors (also referred to as PCA vectors, each N$_{\mathrm{frame}}$ long) and their corresponding eigenvalues. We use the eigenvalues to rank the PCA vectors in an order representing the amount of information encoded by them. Next, we append a unit vector to the set of PCA vectors, and perform a multi-linear regression on the original datacubes (before standardization) using a chosen number of PCA vectors from their ordered set (\Npca{}), which yields a `fit' to the dominant contribution from stellar and telluric systematics in a datacube. We then divide the original datacube by the fit, which yields the detrended datacube. We mask out any residual noisy columns in the detrended datacube in the same way as described in Section 2.2.1 of \cite{panwar_mystery_2024}. An example of this process, which we repeat for each order, is shown in Figure \ref{fig:pca_detrending}. 

The choice of \Npca{} decides how aggressive the detrending is done, and using a high \Npca{} runs the risk of removing part or whole of the planetary signal (e.g. \cite{cheverall_robustness_2023}). For the previously published IGRINS datasets from this program by \cite{brogi_roasting_2023, smith_roasting_2024}, it has been found that fixing \Npca{} to 4 to 6 performs well in terms of removing any visual telluric features, and the results are largely insensitive to the choice of \Npca{}. After testing a range of values from 3 to 12, we find that beyond \Npca{} = 6, the strength of the planetary signal (in terms of CCF S/N and log-likelihood derived confidence intervals, see Section \ref{sec:cross_correlation_analysis}) recovered is not strongly dependent on the choice of \Npca{}. Hence, we fix to \Npca{} = 6 for the remainder of this paper. 

\section{Search for molecular and atomic species using cross-correlation}
\label{sec:cross_correlation_analysis}

We start with performing a search for the planetary signal by computing 2D cross-correlation maps (\KpVsys{} maps) (e.g. \cite{brogi_signature_2012, de_kok_detection_2013}) using a model emission spectrum for the planet's atmosphere. We first describe the calculation of the model planetary emission spectrum, followed by steps to search for the signal in the data using cross-correlation.

We use the \texttt{ScCHIMERA} framework (as used in \cite{arcangeli_h-_2018, piskorz_ground-_2018, mansfield_hstwfc3_2018, line_solar_2021}) to compute average day side pressure temperature (\PT{}) profiles and abundance (gas volume mixing ratio (VMR)) profiles corresponding to 1D radiative convective thermal equilibrium (1D-RCTE) for solar composition (C/O = 0.55, \logZpByZs{} = 0), heat redistribution parameter \textit{f} = 1.98 (based on the trend measured by \cite{parmentier_cloudy_2021}), and planet parameters listed in Table \ref{tab:system_params}. There are two set of planet parameters from the two discovery papers (\cite{rodriguez_kelt-14b_2016, turner_wasp-120_2016}), most of which are consistent with each other within 1$\sigma$, except the planetary radius and mass. We note that the differences arise mainly from the significant differences in stellar radius and mass measured by the two works due to differences in their stellar spectral analysis. In this work, we choose to use both stellar and planetary parameters from \cite{rodriguez_kelt-14b_2016} to be consistent, but we also checked and found that using the \cite{turner_wasp-120_2016} parameters do not change our final results significantly.

\begin{figure*}
\centering
\includegraphics[width=0.9\textwidth]{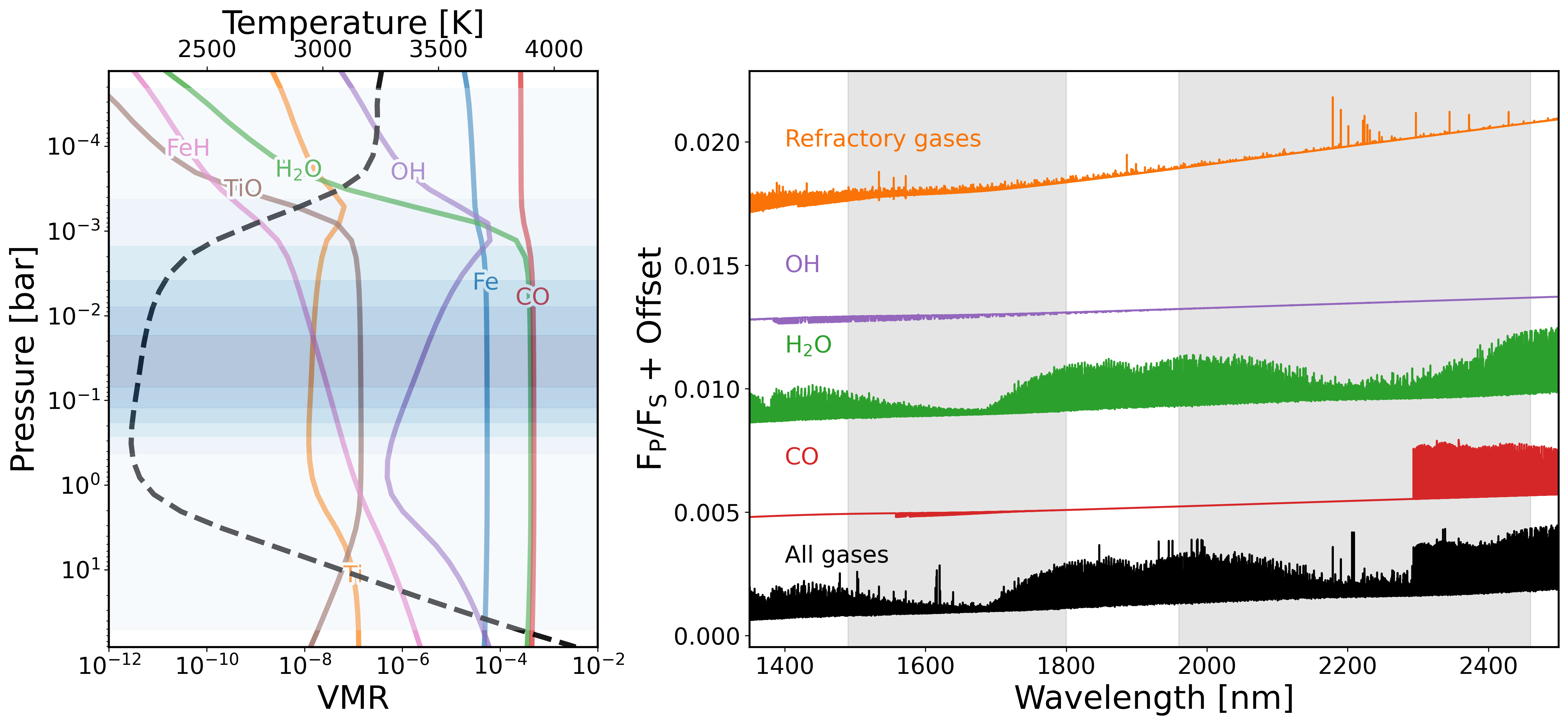}
\caption{(Left panel) Vertical abundance (volume-mixing ratio (VMR)) profiles (solid lines) and \PT{} profile (dashed line) and the emission spectrum from all and individual molecular and atomic species as predicted from 1D radiative-convective thermal equilibrium (1D-RCTE) for WASP-122b. The shaded region in the left panel shows the wavelength integrated contribution function (\protect\cite{knutson_multiwavelength_2008}) calculated using \texttt{GENESIS} taking the VMR and \PT{} profiles from the \texttt{ScCHIMERA} 1D-RCTE model as input. (Right panel) Emission spectrum (ratio of planet flux F$_\mathrm{P}$ to the stellar flux F$_\mathrm{S}$) corresponding to the 1D-RCTE VMR and \PT{} profiles calculated using \texttt{GENESIS} for all the species, and from selected individual molecular and refractory species, vertically shifted for clarity. C and O bearing species \CO{}, \HtwoO{}, \COtwo{} and \OH{}, and refractory bearing species (Fe I, Ti I, TiO, FeH) have detectable features in the IGRINS wavelength coverage (shaded region).}
\label{fig:1D_RCTE_model}
\end{figure*}

We use the \PT{} and VMR profiles from the 1D-RCTE model as inputs to {\tt GENESIS} \citep{gandhi_genesis_2017, pinhas_h2o_2019} to compute the 1D-RCTE planetary spectrum (\Fp{}) at a constant resolution R ($\lambda$/$\Delta\lambda$) = 250000. We include continuum opacities from collision induced absorption (H$_2$-H$_2$ and H$_2$-He), and H$^{-}$ bound-free and H-e$^{-}$ free-free absorption. We also include gas opacities from those atomic and molecular species which have predicted VMR from 1D-RCTE models is $>10^{-7}$, have high cross-sections in the IGRINS wavelength range (taking values at 2000 K, 1 bar for reference), and for which reliably accurate and complete high-resolution opacity data are available. These include molecular, volatile and refractory bearing species: \CO{} \citep{rothman_hitemp_2010, li_rovibrational_2015}, \HtwoO{} \citep{polyansky_exomol_2018}, \COtwo{} \citep{huang_semi-empirical_2013, huang__ames-2016_2017}, \OH{} \citep{buldyreva_simple_2022}, TiO \citep{mckemmish_exomol_2019}, FeH \citep{bernath_mollist_2020}, MgO \citep{buldyreva_simple_2022}, Fe I and Ti I \citep{kurucz_including_2018}. We then rotationally broaden the \Fp{} by convolving it with a rotation kernel corresponding to the projected line of sight planetary rotational velocity ($v_\mathrm{P}~\sin~i$, assuming the planet to be tidally locked). 

For the stellar spectrum (\Fs{}), we take the high-resolution PHOENIX spectrum \cite{husser_new_2013}, broaden it by a rotational kernel corresponding to the projected line of sight stellar rotational velocity ($v_\mathrm{S}~\sin~i$), and smooth it using a Gaussian kernel with standard deviation of 200 elements. We also convolve both \Fp{} and \Fs{} to the instrumental resolution of IGRINS, assuming a Gaussian kernel for the instrument's line-spread function with the full-width half-maxima (FWHM) equivalent to the ratio of the model resolution (R = 250000) to the instrument resolution (R = 45000). We assume the IGRINS resolution to be constant across the wavelength range. 

The planet to star contrast \FpFs{} (hereafter referred to as the model spectrum), which we eventually use to perform the cross-correlation, is computed as the ratio of rotationally and instrumentally broadened \Fp{} and \Fs{} (including scaling by the ratio of planet-to-stellar disk area ((R$_\mathrm{P}$/R$_\mathrm{S}$)$^2$)). The 1D-RCTE \PT{} and VMR profiles, and the corresponding planet to star contrast \FpFs{} from all species and individual molecular and refractory species are shown in Figure \ref{fig:1D_RCTE_model}. The 1D-RCTE predicts an inverted \PT{} profile, with a steep inversion at pressures lower than 0.001 bar, and weaker inversion at higher pressure ranges of 0.5 to 0.15 bar where the peak contribution to the model spectrum comes from based on the wavelength integrated contribution function \citep{griffith_dusty_1998, knutson_multiwavelength_2008}. The model spectrum as a consequence shows spectral features in emission, with dominant contribution from CO, \HtwoO{}, and refractory species. 

We follow the conventional steps common in literature and also described in more detail in \cite{panwar_mystery_2024} to compute the \KpVsys{} map for the cross-correlation function (CCF), which we describe here in brief along with some differences. Before cross-correlating the model spectrum, we describe the steps we follow to first construct the time and wavelength dependent model spectrum cube \FpFs{}($\phi$, $\lambda$) that accounts for the Doppler shift of the stellar and planetary spectrum in the observer's rest frame.  
The line-of-sight velocity of the planet in the rest frame of the observer, assuming a circular orbit, can be described as:

\begin{equation}
\label{eq:rv_planet}
\mathrm{V_{P}}(\phi) = K_\mathrm{P} \mathrm{sin} (2\pi \phi) + V_\mathrm{sys} + V_\mathrm{bary}(\phi) 
\end{equation}  

whereas the line-of-sight velocity for the star is:
\begin{equation}
\label{eq:rv_star}
\mathrm{V_{S}}(\phi) = V_\mathrm{sys} + V_\mathrm{bary}(\phi) + K_S
\end{equation}  

where $\phi$ is the planetary orbital phase which is a function of time, \Kp{} is the Keplerian velocity semi-amplitude of the planet, \Vsys{} is the radial velocity of the planet and star system with respect to the barycentre of the solar system, K$_{S}$ is the Keplerian velocity of the star, and $V_\mathrm{bary}(\phi)$ is the barycentric earth radial velocity (BERV) with respect to the observatory's rest frame. For a hot Jupiter like WASP-122b, K$_{S}$ is of the order of $\sim$100 m/s, which is negligible and quasi-stationary in the comparison to the planetary velocity $\sim$100 km/s. We construct \FpFs($\phi$, $\lambda$) by Doppler shifting \Fp($\lambda$) and \Fs($\lambda$) by V$_{\mathrm{P}}(\phi)$ and V$_{\mathrm{S}}(\phi)$ respectively (Equations \ref{eq:rv_planet} and \ref{eq:rv_star}) and taking their ratio (calculation of \Fp($\lambda$) and \Fs($\lambda$) includes scaling by (R$_\mathrm{P}$/R$_\mathrm{S}$)$^2$). To emulate the effect of PCA detrending on the planetary signal in the data, we perform model reprocessing on \FpFs($\phi$, $\lambda$) using the same steps as described in \cite{panwar_mystery_2024}. This involves first creating model injected data cube F$_{\mathrm{data + model}}$ by injecting \FpFs($\phi$, $\lambda$) into the original data cube F$_{\mathrm{data}}$ (before PCA detrending) for a given pair of \Kp and \Vsys values:

\begin{equation}
\label{eq:inj_model_rep}
\mathrm{F_{data + model}}(\phi, \lambda) = \mathrm{F_{data}}(\phi, \lambda) (1 + \mathrm{F_P/F_S}(\phi, \lambda))
\end{equation} 

We then detrend F$_{\mathrm{data + model}}$ using exactly the same set of PCA eigenvectors and steps used to detrend F$_{\mathrm{data}}$, and obtain the reprocessed model\FpFs{}($\phi$, $\lambda$)$_{\mathrm{reprocess}}$ as:

\begin{equation}
\begin{split}
\label{eq:comp_model_rep}
\mathrm{F_P/F_S}(\phi, \lambda)_{\mathrm{reprocess}} = \mathrm{F_{data+model, detrend}}(\phi, \lambda) - \mathrm{F_{data, detrend}}(\phi, \lambda)
\end{split}
\end{equation} 

\begin{figure*}
\centering
\includegraphics[width=1.0\textwidth]{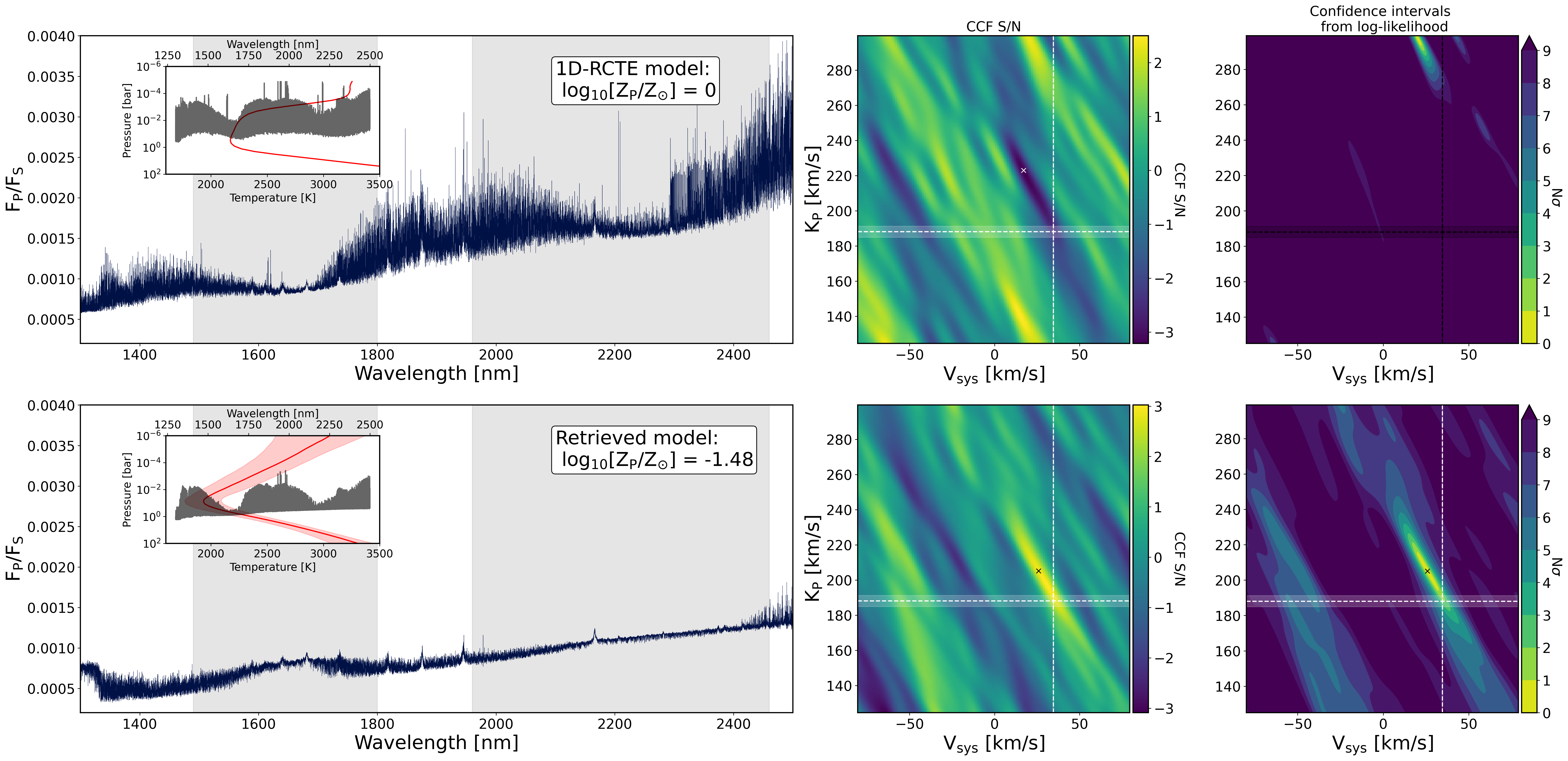}
\caption{Model emission spectrum (\FpFs{}) and the corresponding \KpVsys{} maps for the CCF and log-likelihood derived confidence intervals for the 1D-RCTE model (top row) and the retrieved model (bottom row). In the inset are shown the \PT{} profiles for both cases, with the wavelength dependent photosphere (pressures corresponding to $\tau$ = 2/3 for a given wavelength) overplotted in grey along duplicate X-axis. The solar composition 1D-RCTE model has the photosphere spanning mostly the inverted part of the \PT{} profile, causing all spectral lines to be in emission. In contrast, the retrieved model with $\sim$1.5 dex sub-solar metallicity has a deeper photosphere spanning both non-inverted and inverted region of the \PT{} profile, causing the spectral lines to be both in absorption (bluer end of the spectrum) and emission (redder end of the spectrum). The best fit retrieved model shows a 3$\sigma$ peak in the CCF S/N \KpVsys{} map (bottom panel, second column), and the log-likelihood derived confidence interval map (bottom panel, third column) show a clear detection with the 3$\sigma$ interval tightly constrained but the peak of the signal (marked by 'x') shifted from the expected \Kp{} and \Vsys{} values (marked by dashed lines).}
\label{fig:KpVsys_maps}
\end{figure*}

We compute the CCF \KpVsys{} maps per order by computing the cross-correlation (cross-covariance normalized by the variances) between the reprocessed model spectrum \FpFs{}($\phi$, $\lambda$)$_{\mathrm{reprocess}}$ and the data cube frame by frame (i.e. for each $\phi$) for a grid of \Kp{} and \Vsys{} values with step size of 1 \kms{}. We sum the CCF across all frames to obtain the map for each order, which we then sum across all orders to obtain the total \KpVsys{} map. We compute the signal-to-noise of the peak by first sigma-clipping any outliers in the map beyond 3$\sigma$ threshold, followed by subtracting the median of all the values outside the $\pm$10~\kms{} of the peak within which the signal is localized. We then normalize the median subtracted map by the standard-deviation of all values outside the $\pm$10\kms{} of the peak, resulting in the map for CCF S/N shown in Figure \ref{fig:KpVsys_maps}. We also use the CCF to log-likelihood framework introduced by \cite{brogi_retrieving_2019} to construct the corresponding log-likelihood derived confidence interval map. This is an alternative way to assess the significance of CCF by computing the confidence intervals for the map using the log-likelihood and assuming 2 degrees of freedom corresponding to the two velocities \Kp{} and \Vsys{} (see \cite{panwar_mystery_2024} for more details of the steps). The resulting CCF S/N and confidence interval \KpVsys{} map for the 1D-RCTE model is shown in Figure \ref{fig:KpVsys_maps}. 

\begin{figure*}
\centering
\includegraphics[width=1.0\textwidth]{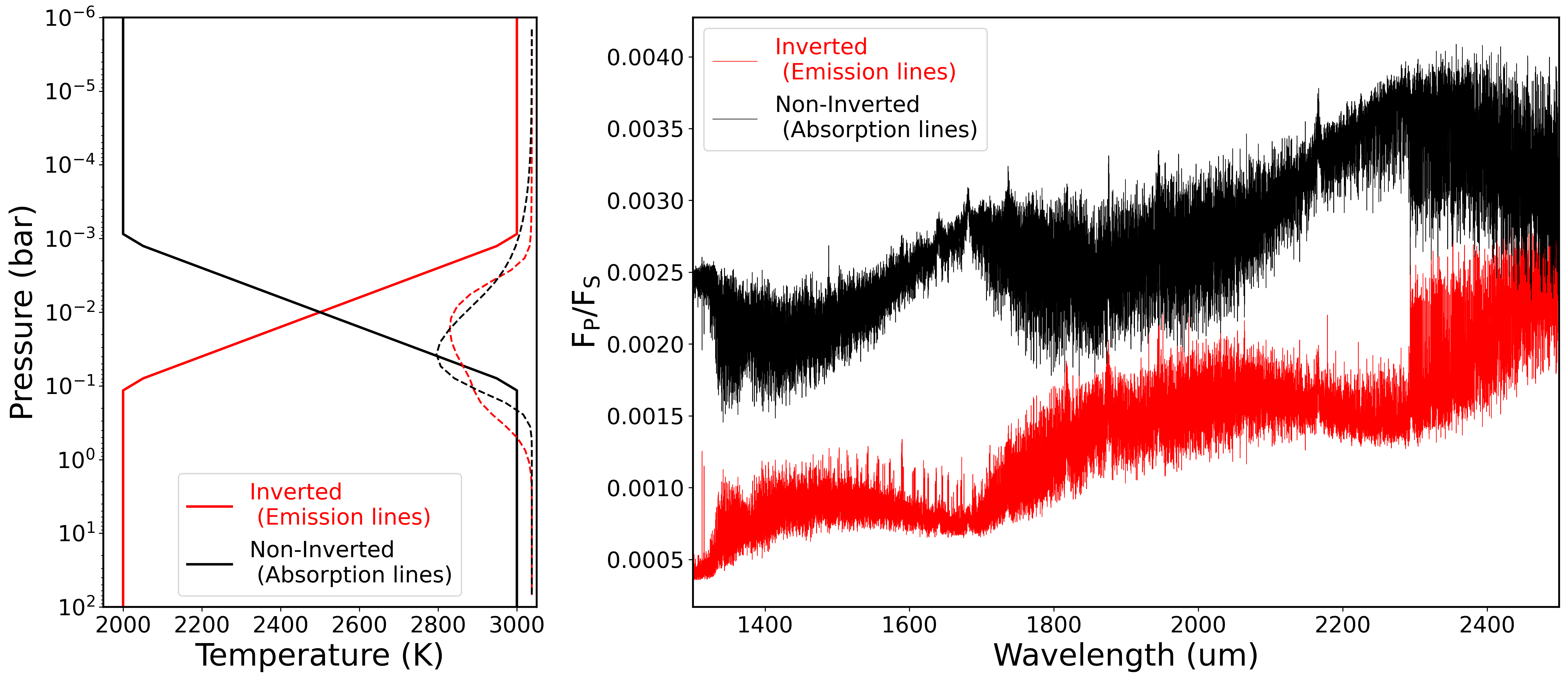}
\caption{Dichotomy between inverted and non-inverted \PT{} profiles (left panel) and the shape of lines in the corresponding spectra (right panel), both assuming solar composition, with an arbitrary vertical offset between them for clarity. The solid lines in the left panel show two fully inverted (red) and fully non-inverted (black) \PT{} profiles prescribed as a linear temperature gradient between two points in the atmosphere. Their respective wavelength integrated contribution functions, assuming solar composition for both, are shown in dashed curve. The inverted profile leads to all spectral lines in emission (right panel), whereas the non-inverted profile yields all spectral lines in absorption.}
\label{fig:inverted_vs_non-inverted}
\end{figure*}

On inspecting the CCF S/N map for the 1D-RCTE model, we find that instead of a positive peak in the CCF at the expected \Kp{} and \Vsys{} there is an anticorrelated peak with S/N $\sim$3$\sigma$ offset by $\sim$ +35 \kms{} and -17 \kms{} in \Kp{} and \Vsys{} respectively. This indicates that there is a mismatch in the shape and contrast of the spectral lines in the solar composition (C/O = 0.55, log$_{10}$Z = 0) 1D-RCTE model spectrum and the planetary signal present in the data. The 1D-RCTE model spectrum has an inverted profile, with all the spectral lines in emission, whereas it is likely that the spectral lines in the data are either all or partly in absorption, which could be causing the anticorrelated signal in the CCF map. We illustrate this dichotomy in Figure \ref{fig:inverted_vs_non-inverted} by computing spectra from solely inverted and non-inverted \PT{} profiles for WASP-122b, assuming a simple linear gradient in temperature with pressure and solar composition. A solely inverted profile shows all spectral lines in emission, whereas solely non-inverted profile shows all lines in absorption. 
The anticorrelated peak in the \KpVsys{} map implies that a solar composition 1D-RCTE model is an incorrect description of the atmosphere, and there are variations in the \PT{} profile and atmospheric abundance that need to be investigated to figure out the scenario that best matches the planetary signal in the data. In the next section, we do this by performing an atmospheric retrieval, which searches across models with a range of atmospheric abundances and thermal structures, to find the best fit atmospheric model in the context of the data, along with statistical constraints on the model parameters.     

%%%%% Figure showing the inversion/non-inversion - absorption/emission dichotomy

\section{Retrieving the Atmospheric Composition and Thermal Structure}
\label{sec:retrieval_analysis}
We move towards obtaining statistical constraints on the composition and thermal structure of WASP-122b's dayside by using the CCF to log-likelihood mapping framework from \cite{brogi_framework_2017} to perform Bayesian inference (commonly referred to as an atmospheric retrieval). This involves parametrizing the forward model spectrum calculated using \texttt{GENESIS} (as described in Section \ref{sec:cross_correlation_analysis}) by parametrizing the Doppler shift (through \Kp{} and \Vsys{}), and the gas VMR and \PT{} profiles. 

We follow the two common methods to parametrize the gas VMR profiles in the literature: 1) assuming free chemistry (e.g. \cite{line_systematic_2013}), where the VMR of each gas is assumed to be vertically constant and is a free parameter, and 2) assuming equilibrium chemistry (e.g. \cite{brogi_roasting_2023, smith_roasting_2024}), where the VMR profile of each gas is calculated assuming equilibrium chemistry using \texttt{FastChem} \citep{stock_fastchem_2018, stock_fastchem_2022}. \texttt{FastChem} takes inputs in the form of \PT{} profile, metallicity (log$_{10}$(Z$_\mathrm{P}$/Z$_{\odot}$), Z$_\mathrm{P}$ is the planet metallicity, Z$_{\odot}$ is the solar metallicity), and C/O ratio, and calculates abundance profiles for chemical species accounting for dissociation, ionization, and condensation. For each \texttt{FastChem} calculation, we use the value of the planet metallicity relative to solar composition (log$_{10}$(Z$_\mathrm{P}$/Z$_{\odot}$) to scale the abundances of all elements first, and then use the value of C/O to scale the abundance of Carbon relative to Oxygen.

We employ a slight variation to the free chemistry setup by considering the VMR of all species constant with pressure except \HtwoO{}. For \HtwoO{}, we expect significant thermal dissociation onset around 1 mbar as indicated by the RCTE model (as shown in Figure \ref{fig:1D_RCTE_model}). Hence, we use the empirical power law coefficients presented by \cite{parmentier_thermal_2018} to prescribe the \HtwoO{} VMR profile accounting for thermal dissociation, fixing all the parameters except the deep abundance of \HtwoO{} which we fit for in the retrieval. 

For both free and equilibrium chemistry retrievals, we follow the same setup for \texttt{GENESIS} as described in Section \ref{sec:cross_correlation_analysis} to compute the forward model spectrum cube \FpFs{}($\phi$, $\lambda$) along with model reprocessing, and considering contributions from the following molecules: CO, \HtwoO{}, \OH{}, \HtwoS{}, TiO, FeH, MgO, and atomic species: Fe and Ti. We also include continuum opacities from collision induced absorption (H$_2$-H$_2$ and H$_2$-He), and H$^{-}$ bound-free and H-e$^{-}$ free-free absorption. We choose to use the 6 parameter B{\'e}zier spline parametrization for the \PT{} profile, as introduced by \cite{smith_roasting_2024}. This involves defining four points or nodes: (log$_{10}$P$_0$, T$_0$), (log$_{10}$P$_1$, T$_1$), (log$_{10}$P$_2$, T$_2$), (log$_{10}$P$_3$, T$_3$); log$_{10}$P$_0$ > log$_{10}$P$_1$ > log$_{10}$P$_2$ > log$_{10}$P$_3$, in the pressure-temperature space and using them as control points to compute a B{\'e}zier curve, which is then interpolated onto the pressure grid of \texttt{GENESIS}, which in this work we consider as 50 layers spaced linearly in log-space between 10$^{-6}$ to 10$^2$ bar. We fix the pressure values for two nodes to the bottom (log$_{10}$P$_0$ [bar] = 2) and top of the atmosphere (log$_{10}$P$_3$ [bar] = -6), which leaves 6 free parameters:  T$_0$, log$_{10}$P$_1$, T$_1$, log$_{10}$P$_2$, T$_2$, T$_3$. This B{\'e}zier spline parametrization provides a balance between the flexibility in the shape of the profile and the number of free parameters needed to describe it. It allows for both a monotonically inverted or non-inverted \PT{} profile, and a mix of both in the lower and upper of the atmospheres. The latter is especially relevant in the case of WASP-122b, as we discuss in more detail in Section \ref{sec:discuss:met_TP}.  

The total set of free parameters and their priors for both the equilibrium and free chemistry retrievals are shown in Table \ref{tab:eq_params_fit} and \ref{tab:free_params_fit} respectively. We run the retrieval using our code \texttt{crocodel}\footnote{https://github.com/vatsalpanwar/crocodel} which couples the parametrized log-likelihood calculation with the nested sampling algorithm {\tt MultiNest} \citep{feroz_multinest_2009} implemented as {\tt pymultinest} \citep{buchner_x-ray_2014} to compute the marginalized posterior distributions for the free model parameters. We run {\tt MultiNest} with the number of live points set to 1000, and a tolerance on the Bayesian evidence (Z) as $\Delta$lnZ = 0.5. The retrieval converges when the tolerance on $\Delta$lnZ is reached, with $\sim$ 100000 log-likelihood evaluations, resulting in 8822 and 11934 equal weighted posterior samples for the free and equilibrium chemistry cases respectively.

%%%% Tables of priors and best fit parameters goes here. Make one table for equilibrium chemistry retrieval, and one for free chemistry retrieval
We summarize the results from the equilibrium and free chemistry retrievals in the brief corner plots shown in Figures \ref{fig:brief_retreival_cornerplot_eq_chem} and \ref{fig:brief_retreival_cornerplot_free_chem} respectively. The full corner plots for all the parameters are shown in Figures \ref{fig:full_eq_chem_retrieval_posteriors} and \ref{fig:full_free_chem_retrieval_posteriors}. Table \ref{tab:eq_params_fit} and \ref{tab:free_params_fit} list the best fit and $±1\sigma$ constraints obtained for each free parameter from the two retrievals. We also tested the sensitivity of retrievals to the choice of \Npca{}, by repeating the equilibrium chemistry retrieval for a range of values for \Npca{} = (4, 5, 6, 7, 8). We find that the constraints on the abundances and the \PT{} profile obtained from the retrievals with varying \Npca{} are consistent with each other within 1$\sigma$, whereas the velocities (\Kp{}, and \Vsys{}) are consistent within 2$\sigma$.

%%%%%%%%%% Equilibrium chemistry
\begin{table}
\centering
\renewcommand{\arraystretch}{1.5} %
\caption{Best fit and 1$\sigma$ values for the parameters from equilibrium chemistry retrieval, assuming 4 node B{\'e}zier spline \PT{} parametrization.}
\begin{tabular}{|c|c|c|}
\hline
Parameter & Prior Bounds & Best fit $\pm$ 1$\sigma$ \\
\hline
K$_{\mathrm{P}}$ (km s$^{-1}$) & [160.00, 250.00] & 205.16$_{-7.61}^{+7.35}$ \\
V$_{\mathrm{sys}}$ (km s$^{-1}$) & [5.00, 50.00] & 25.86$_{-4.24}^{+4.49}$ \\
T$_{0}$ [K] & [400.00, 3500.00] & 3315.36$_{-248.52}^{+132.87}$ \\
T$_{1}$ [K] & [400.00, 3500.00] & 1402.91$_{-699.01}^{+936.04}$ \\
log$_{10}$P$_{1}$ [bar] & [-6.00, 2.00] & -2.58$_{-1.09}^{+0.92}$ \\
T$_{2}$ [K] & [400.00, 3500.00] & 1532.01$_{-730.66}^{+979.88}$ \\
log$_{10}$P$_{2}$ [bar] & [-6.00, 2.00] & 1.44$_{-0.86}^{+0.41}$ \\
T$_{3}$ [K] & [400.00, 3500.00] & 3069.83$_{-482.52}^{+299.21}$ \\
log$_{10}$[Z$_{\mathrm{P}}$/Z$_{\odot}$] & [-6.00, 2.00] & -1.48$_{-0.22}^{+0.26}$ \\
C/O & [0.00, 2.00] & 0.36$_{-0.22}^{+0.23}$ \\
\hline
\end{tabular}
\label{tab:eq_params_fit}
\end{table}

%%%%%%%%%% Equilibrium chemistry with Linear TP
\begin{table}
\centering
\renewcommand{\arraystretch}{1.5} %
\caption{Best fit and 1$\sigma$ values for the parameters from equilibrium chemistry retrieval, assuming linear gradient \PT{} with no inversion.}
\begin{tabular}{|c|c|c|}
\hline
Parameter & Prior Bounds & Best fit $\pm$ 1$\sigma$ \\
\hline
K$_{\mathrm{P}}$ (km s$^{-1}$) & [160.00, 250.00] & 205.57$_{-7.65}^{+7.86}$ \\
V$_{\mathrm{sys}}$ (km s$^{-1}$) & [5.00, 50.00] & 25.80$_{-4.40}^{+4.27}$ \\
log$_{10}$P$_{1}$ [bar] & [-1.00, 7.00] & 0.55$_{--4.59}^{+5.45}$ \\
T$_{1}$ [K] & [400.00, 3500.00] & 3044.69$_{-368.34}^{+320.58}$ \\
log$_{10}$P$_{2}$ [bar] & [-1.00, 7.00] & -0.31$_{--4.83}^{+5.16}$ \\
T$_{2}$ [K] & [400.00, 3500.00] & 1975.39$_{-205.74}^{+196.09}$ \\
log$_{10}$[Z$_{\mathrm{P}}$/Z$_{\odot}$] & [-6.00, 2.00] & -1.66$_{-0.26}^{+0.27}$ \\
C/O & [0.00, 2.00] & 0.27$_{-0.19}^{+0.24}$ \\
\hline
\end{tabular}
\label{tab:eq_params_fit_Linear_TP}
\end{table}

%%%%%%%%%% Free chemistry
\begin{table}
\centering
\renewcommand{\arraystretch}{1.5} %
\caption{Best fit and 1$\sigma$ values for the parameters from free chemistry retrieval, assuming 4 node B{\'e}zier spline \PT{} parametrization.}
\begin{tabular}{|c|c|c|}
\hline
Parameter & Uniform Prior Bounds & Best fit $\pm$ 1$\sigma$ \\
\hline
K$_{\mathrm{P}}$ (km s$^{-1}$) & [160.00, 250.00] & 204.08$_{-4.92}^{+4.74}$ \\
V$_{\mathrm{sys}}$ (km s$^{-1}$) & [5.00, 50.00] & 26.34$_{-2.74}^{+2.92}$ \\
T$_{0}$ [K] & [400.00, 3500.00] & 3370.89$_{-176.69}^{+94.28}$ \\
T$_{1}$ [K] & [400.00, 3500.00] & 1253.08$_{-596.67}^{+949.41}$ \\
log$_{10}$P$_{1}$ [bar] & [-6.00, 2.00] & -2.92$_{-1.04}^{+0.97}$ \\
T$_{2}$ [K] & [400.00, 3500.00] & 1679.11$_{-740.91}^{+938.82}$ \\
log$_{10}$P$_{2}$ [bar] & [-6.00, 2.00] & 1.59$_{-0.59}^{+0.29}$ \\
T$_{3}$ [K] & [400.00, 3500.00] & 3217.55$_{-286.41}^{+191.18}$ \\
log$_{10}$[CO] & [-15.00, -1.00] & -9.77$_{-3.42}^{+3.65}$ \\
log$_{10}$[H$_{2}$O] & [-15.00, -1.00] & -4.86$_{-0.13}^{+0.14}$ \\
log$_{10}$[OH] & [-15.00, -1.00] & -11.04$_{-2.52}^{+2.54}$ \\
log$_{10}$[FeH] & [-15.00, -1.00] & -11.50$_{-2.21}^{+2.20}$ \\
log$_{10}$[TiO] & [-15.00, -1.00] & -10.94$_{-2.59}^{+2.68}$ \\
log$_{10}$[MgO] & [-15.00, -1.00] & -10.25$_{-3.02}^{+1.24}$ \\
log$_{10}$[Fe] & [-15.00, -1.00] & -2.44$_{-7.21}^{+0.91}$ \\
log$_{10}$[Ti] & [-15.00, -1.00] & -6.68$_{-4.72}^{+0.98}$ \\
log$_{10}$H$^{-}$ & [-15.00, -1.00] & -12.62$_{-1.48}^{+1.49}$ \\
\hline
\end{tabular}
\label{tab:free_params_fit}
\end{table}

%%%%%%%%%%%%%%%%%%%%%%%%%%%%%%%%%%%%%%%%%%%%%%%%%%%%%%%%%%%
\subsection{Constraints on the Abundances and Thermal Structure of WASP-122b}
\label{sec:results:constraints}
%%%%%%%%%%%%%%%%%%%%%%%%%%%%%%%%%%%%%%%%%%%%%%%%%%%%%%%%%%%
%%% Brief eq chem posteriors 
% \begin{figure*}
% \centering
% \includegraphics[width=0.9\textwidth]{figures/eq_chem_posteriors_with_TP_brief.png}
% \caption{Cornerplot showing the 1D and 2D posterior distributions for some of the free parameters constrained by the equilibrium chemistry retrieval (described in Section \ref{sec:retrieval_analysis}). The inset shows the best fit (solid line) and the 1$\sigma$ bounds constrained for the \PT{} profile by the retrieval, with the wavelength dependent $\tau(\lambda)$ = 2/3 photosphere overplotted. See Figure \ref{fig:full_eq_chem_retrieval_posteriors} for the cornerplot including all the free parameters in this retrieval.}
% \label{fig:brief_retreival_cornerplot_eq_chem}
% \end{figure*}
\begin{figure*}
\centering
\includegraphics[width=0.9\textwidth]{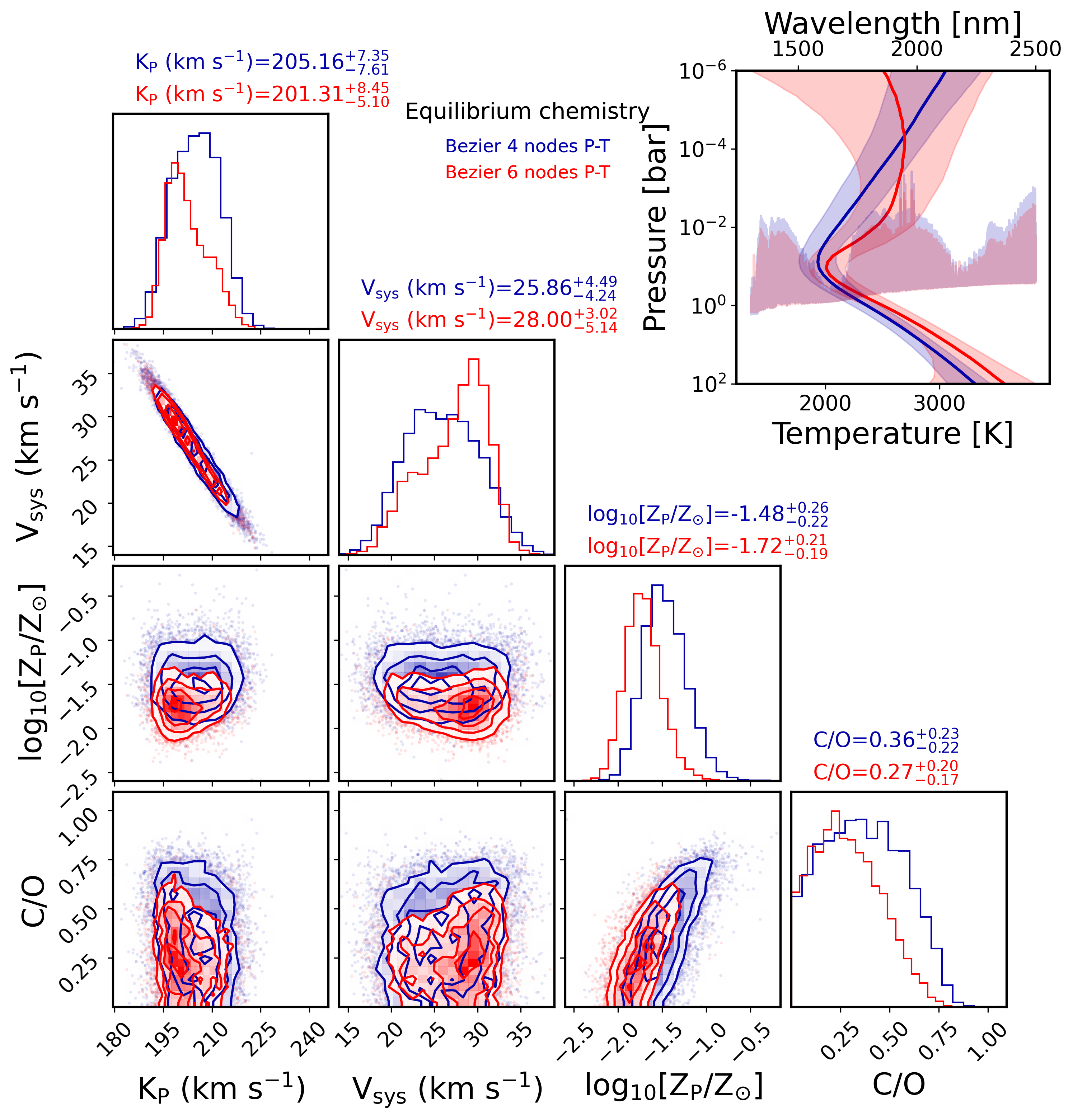}
\caption{Cornerplot showing the 1D and 2D posterior distributions for some of the free parameters constrained by the equilibrium chemistry retrieval (described in Section \ref{sec:retrieval_analysis}) using both 4 node (blue) and 6 node (red) B{\'e}zier spline \PT{} parametrization. The inset in, two corresponding colours, shows the best fit (solid line) and the 1$\sigma$ bounds for the \PT{}, and the wavelength dependent $\tau(\lambda)$ = 2/3 photosphere overplotted in the background. See Figure \ref{fig:full_eq_chem_retrieval_posteriors} for the cornerplot including all the free parameters in this retrieval.}
\label{fig:brief_retreival_cornerplot_eq_chem}
\end{figure*}

%%% Brief eq chem posteriors with Linear TP profile 
\begin{figure*}
\centering
\includegraphics[width=0.9\textwidth]{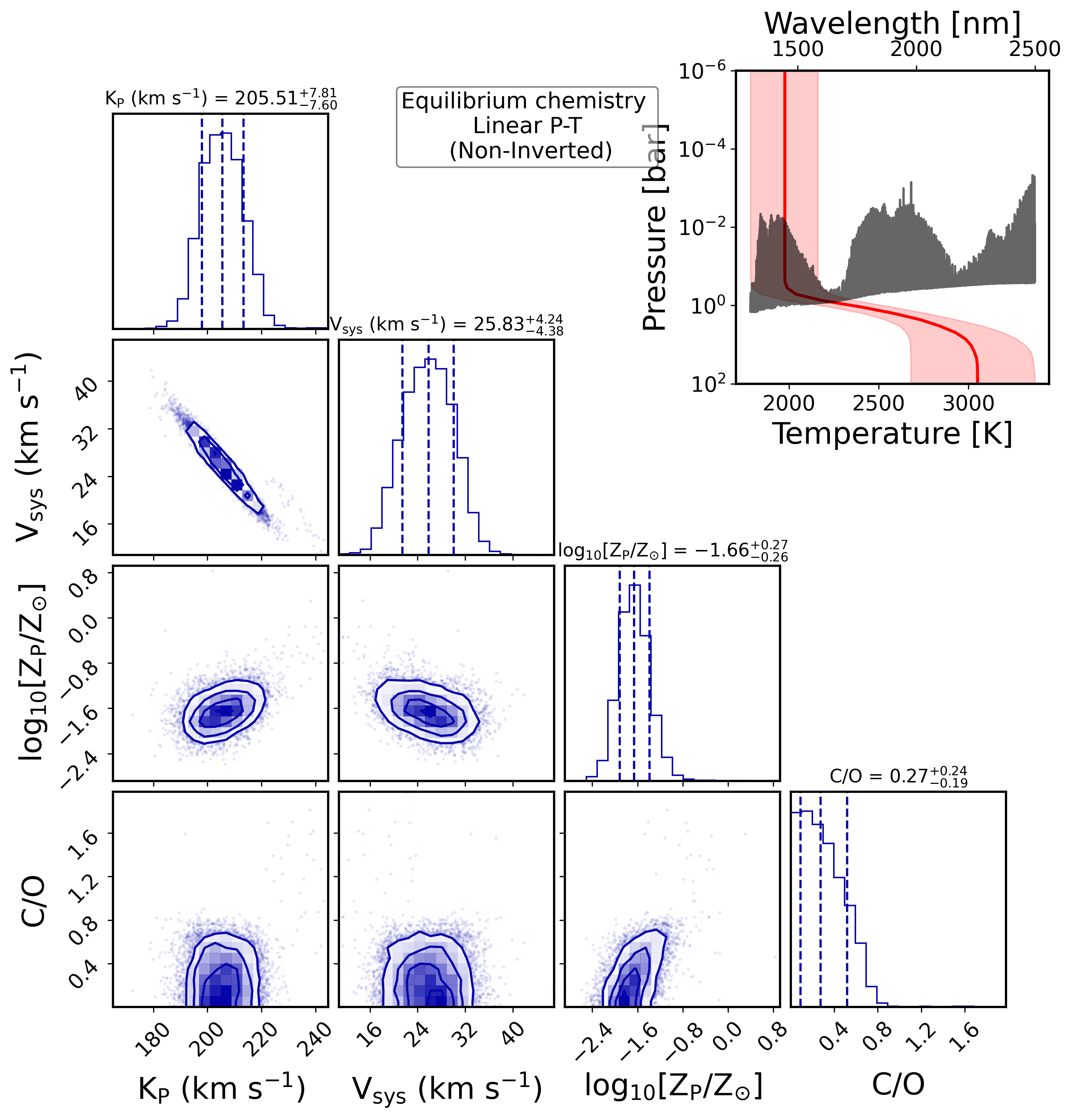}
\caption{Same as Figure \ref{fig:brief_retreival_cornerplot_eq_chem}, but using a linear gradient \PT{} with no inversion.}
\label{fig:brief_retreival_cornerplot_eq_chem_Linear_TP}
\end{figure*}

%%% Brief free chem posteriors 
\begin{figure*}
\centering
\includegraphics[width=0.9\textwidth]{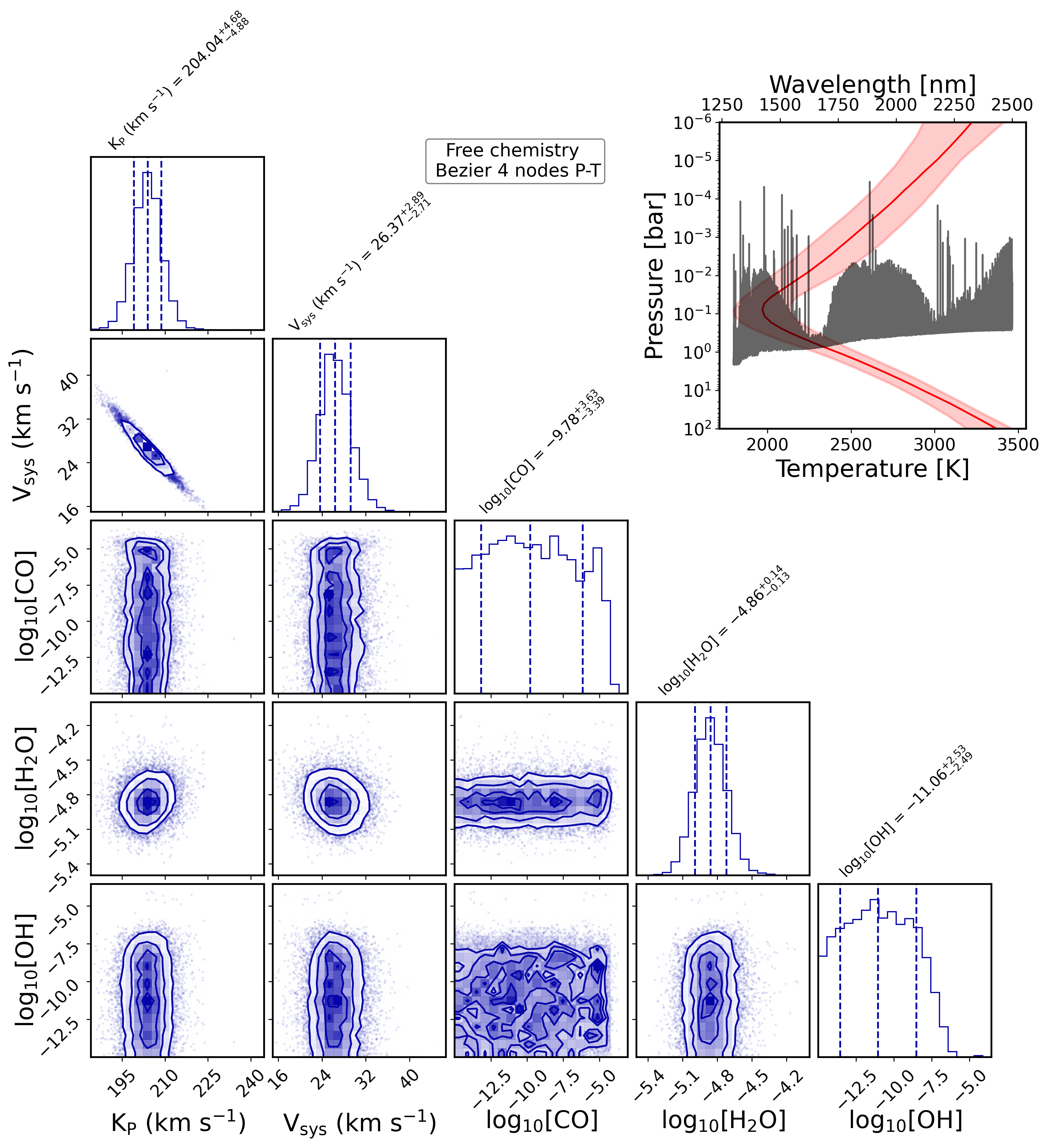}
\caption{Same as Figure \ref{fig:brief_retreival_cornerplot_eq_chem}, but for the free-chemistry retrieval. Note that the \HtwoO{} abundance plotted here is the deep abundance of \HtwoO{}, and all species except \HtwoO{} in the free-chemistry retrieval are assumed to have vertically uniform abundance profile.}
\label{fig:brief_retreival_cornerplot_free_chem}
\end{figure*}

We find that both free and equilibrium chemistry retrievals successfully converge and constrain the \Kp{} and \Vsys{} velocities and the \PT{} profile. In particular, the posteriors from the equilibrium chemistry retrieval (Figure \ref{fig:brief_retreival_cornerplot_eq_chem}) indicates a significantly metal-depleted atmosphere with \logZpByZs{} = $-$1.48$_{-0.22}^{+0.25}$ dex, which translates to 0.033$_{-0.016}^{+0.018}$ $\times$ solar metallicity. We find solar/sub-solar C/O ratio = 0.36$_{-0.22}^{+0.22}$ (3$\sigma$ upper limit 0.82) consistent with solar C/O ratio of 0.55 \cite{asplund_chemical_2021}. This agrees with results from the free chemistry retrieval which finds a $\sim$2 dex sub-solar abundance for \HtwoO{} \logWater{} = -4.86$_{-0.13}^{+0.14}$ with a non-detection of CO, OH (Figure \ref{fig:brief_retreival_cornerplot_free_chem}) and all the refractory species (Figure \ref{fig:full_free_chem_retrieval_posteriors}). 
% From the free chemistry retrieval, we derive \logZpByZs{} = -2.02$_{-0.18}^{+0.14}$ by converting the posteriors for individual species to posteriors for [M/H] (following the same steps described in \citealt{panwar_mystery_2024}). Note that the sub-solar metallicity constrained by the free-chemistry retrieval largely hinges around the subsolar [O/H] ratio constrained through the detection of \HtwoO{} at sub-solar abundance. 

\begin{figure*}
\centering
\includegraphics[width=0.9\textwidth]{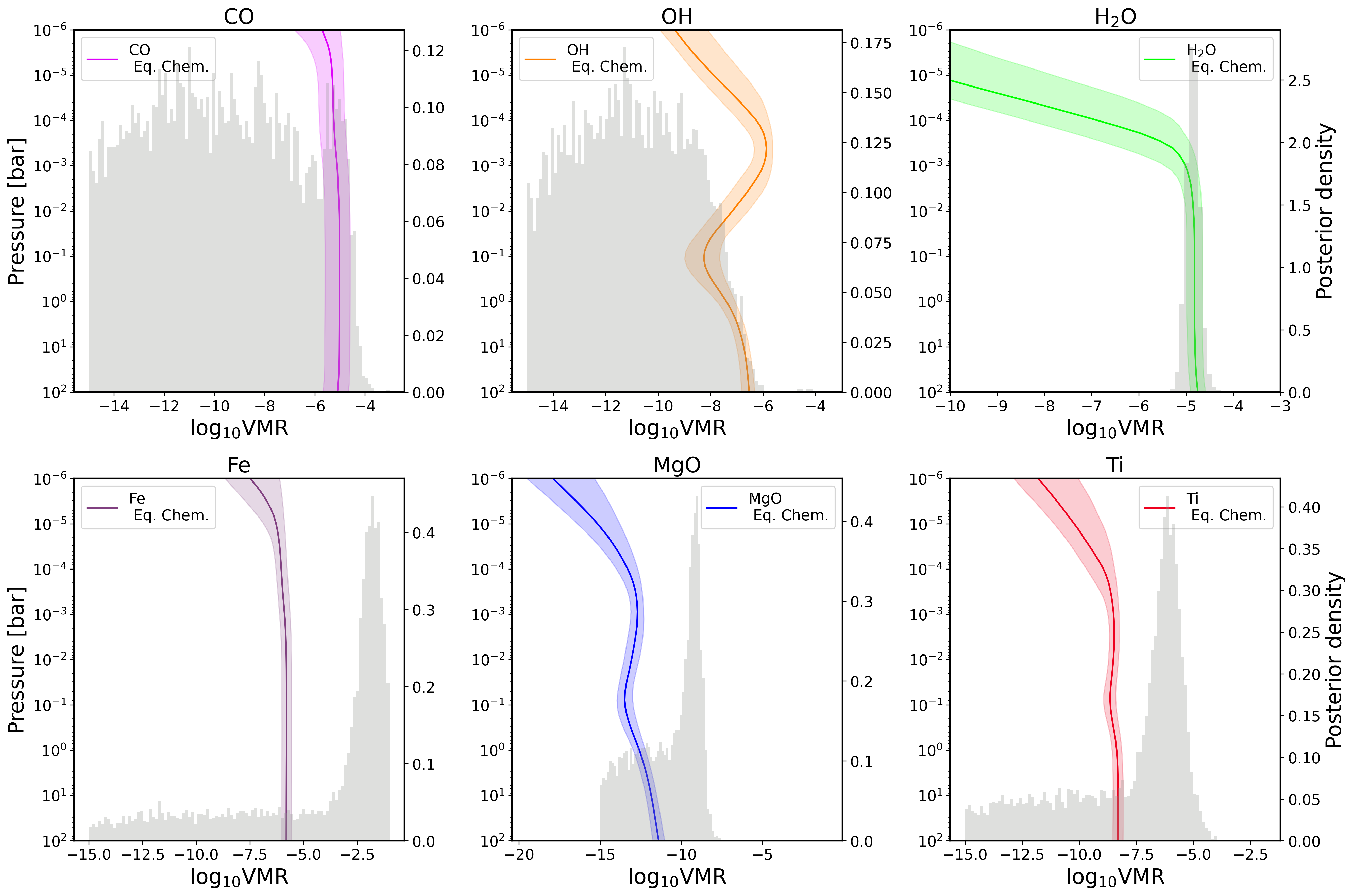}
\caption{Vertical abundance profiles from the equilibrium chemistry retrieval, and posteriors for corresponding species from the free chemistry retrieval assuming vertically uniform abundances (except for \HtwoO{}, see Section \ref{sec:retrieval_analysis}.) By comparing the inferences for the molecular abundances from both equilibrium (which only has metallicity and C/O ratio as the free parameters deciding the abundance profiles at equilibrium) and free-chemistry posteriors, we conclude that \HtwoO{} is the only species for which the data yields a constraint with bounded posteriors. All the other species, despite having a peak in their posteriors, have a finite tail that fills up the retrieval prior, and hence only yield upper limits.}
\label{fig:retrieved_abundances}
\end{figure*}

We find that some refractory and refractory bearing species like Fe, MgO, and Ti show peaks in their posterior distribution (see Figure \ref{fig:full_free_chem_retrieval_posteriors}), but the overall distribution has a finite tail towards lower abundances up to the lower bound of the prior, which indicates that these species are unconstrained by the retrieval. We also find that the vertical abundance profiles constrained from the equilibrium chemistry retrieval are consistent with the constraints on the vertically uniform abundances from the free chemistry retrieval, as shown in Figure \ref{fig:retrieved_abundances}. We also ran a free chemistry retrieval with the same setup but without Fe and Ti and find negligible support for model with Fe and Ti (Bayes factor based on Bayesian evidence (Z) between the two models is $\Delta$lnZ $\sim$ 1).

%%% Brief free chem posteriors 

Similar to the steps followed in previous studies (e.g. \cite{panwar_mystery_2024, brogi_roasting_2023}) we also derived the posteriors for atmospheric metallicity \logZpByZs{} from the free chemistry retrieval as shown in Figure \ref{fig:1D_C_to_O_and_metallicity}. We considered only the posterior samples with CO abundance above $-6.5$ dex, because for smaller values of CO abundance the derived posterior for C/O gets heavily piled up at values close to 0, making it difficult to estimate the upper limit. We find that the derived posteriors from the free chemistry retrieval indicate the \logZpByZs{} = $-$1.78$^{+0.34}_{-0.22}$ (driven by planetary O/H measured through detection of \HtwoO{}), and 3$\sigma$ upper limit on C/O = 0.91. This is consistent with the constraints on \logZpByZs{} and the upper limit on the C/O ratio obtained from the equilibrium chemistry retrieval. The Bayes factor between the free and equilibrium chemistry is negligible, implying no strong preference for either ($\Delta$lnZ $\sim$ 0.4).

The \PT{} profile constrained by both the retrievals are also consistent with each other well within 1$\sigma$ (see inset in Figures \ref{fig:brief_retreival_cornerplot_eq_chem} and \ref{fig:brief_retreival_cornerplot_free_chem}). Note that the free-parameters for our \PT{} parametrization are the three pressure and temperature nodes (log$_{10}$P, T) for the B{\'e}zier spline parametrization, and the nodes themselves do not lie on the \PT{} profile but are used as control points to compute the \PT{} profile for a given range of pressures. Hence, while the posteriors for the nodes themselves do not have bounded constraints as seen from their posteriors in Figure \ref{fig:full_free_chem_retrieval_posteriors}, the \PT{} profile itself is well constrained in the range of pressures we compute the emission spectrum using \texttt{GENESIS}. We find that both the free and equilibrium chemistry retrievals indicate a \PT{} profile which going from high to low pressures is non-inverted until around 0.1 bar, where it switches to become inverted. 

To test how robust the shape of the \PT{} profile is, we ran retrievals with alternative parametrizations for the profile. This also checks how sensitive the constrained abundances are to the choice of \PT{} parametrization. We first tested a retrieval with 6 node B{\'e}zier profile for the equilibrium chemistry case, which after fixing the pressure values for the nodes at the top and bottom of the pressure range, adds four more free parameters to the original 4 node B{\'e}zier profile. We find that both the retrieved \PT{} profile and the metallicity are consistent between 4 and 6 node B{\'e}zier spline parametrization (as shown in Figure \ref{fig:brief_retreival_cornerplot_eq_chem} and discussed further in Appendix \ref{app:retrieval_madhu_seager}). Going forward, we refer to the constraints from the 4 node B{\'e}zier profile in the context of equilibrium chemistry retrievals, unless specified otherwise.

\begin{figure}
\centering
\includegraphics[width=0.5\textwidth]{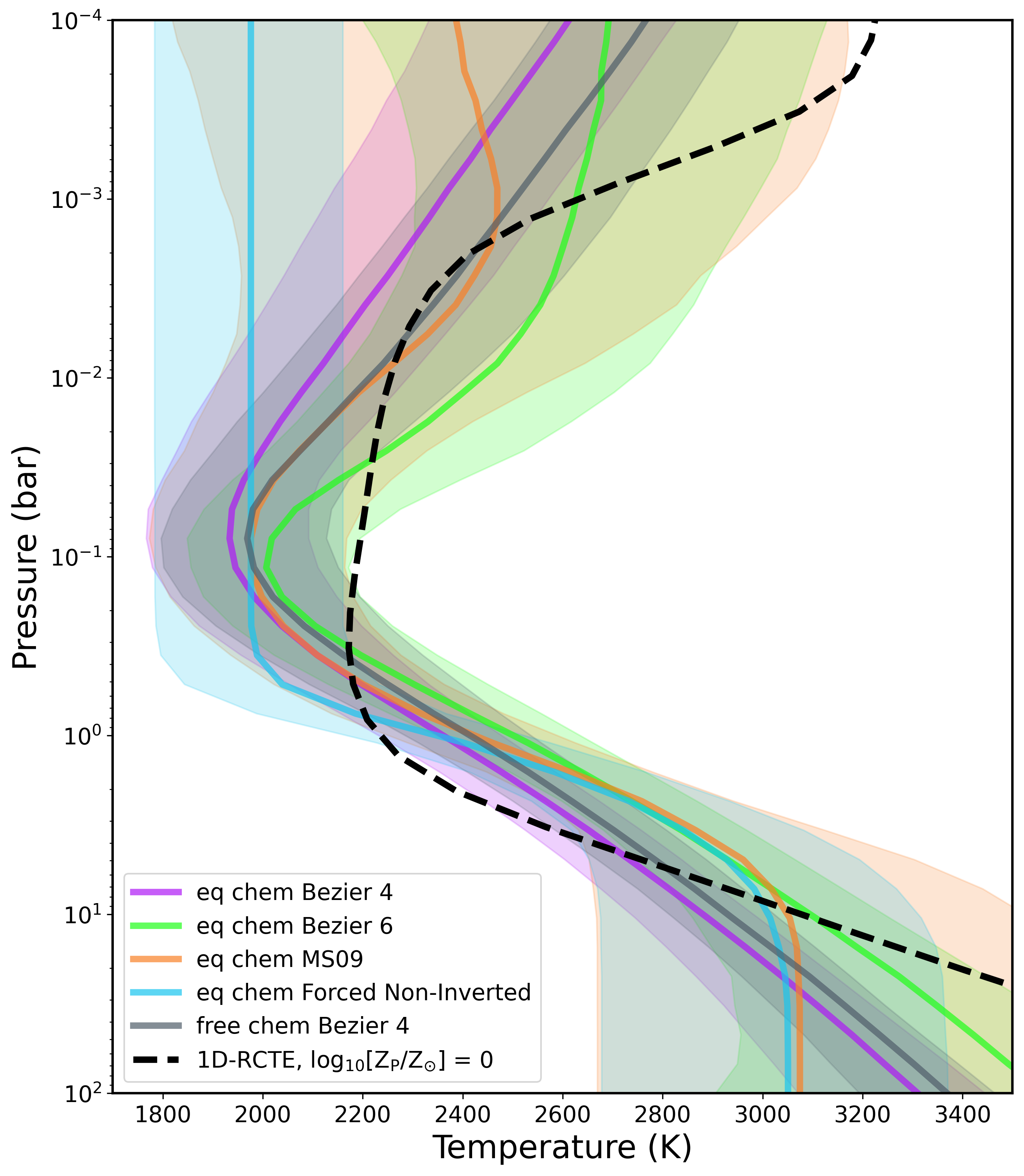}
\caption{Comparison of the \PT{} profiles constrained from a range model parametrizations, and from the 1D-RCTE model assuming solar metallicity (dashed black). The solid lines show the best fit \PT{} profiles obtained from 5 different retrievals, one of which assumes free chemistry and uses the 4 node B{\'e}zier spline profile, and the other four of which assume equilibrium chemistry with a range of \PT{} profile parametrizations: 4 node B{\'e}zier spline, 6 node B{\'e}zier spline, \protect\cite{madhusudhan_temperature_2009} (MS09) profile, and a forced non-inverted linear gradient profile. The shaded region for each profile represent the $\pm$1$\sigma$ intervals indicated by the corresponding retrieval posteriors.}
\label{fig:TP_profiles_comparison}
\end{figure}

We also ran two equilibrium chemistry retrievals using a simpler linear gradient \PT{} profile (as used in \cite{panwar_mystery_2024}), one forced to be only inverted and the other forced to be only non-inverted. The two profiles have a linear gradient in temperature with respect to log(P) between two points in the atmosphere (four free parameters : $P_1$, $T_1$, $P_2$, $T_2$), and is isothermal otherwise. Comparing the Bayesian evidence from \texttt{MultiNest} for these two retrievals, we find that the non-inverted \PT{} profile is significantly preferred over the inverted one by $\Delta$lnZ = 9. The retrieval with non-inverted profile converges to parameters consistent with the B{\'e}zier spline profile retrieval within 1$\sigma$, as shown in Figure \ref{fig:brief_retreival_cornerplot_eq_chem_Linear_TP}, and Table \ref{tab:eq_params_fit_Linear_TP}. However, we also find that the $\Delta$lnZ between the non-inverted linear \PT{} profile and B{\'e}zier spline profile both 4 and 6 nodes is negligible, indicating that there is no strong preference for the switch to inversion indicated by the B{\'e}zier spline profile. We also confirm this by running another equilibrium chemistry retrieval using the \PT{} profile parametrization from \cite{madhusudhan_temperature_2009}, as shown and discussed in more detail in Appendix \ref{app:retrieval_madhu_seager}. The \PT{} profiles obtained from different model parametrizations are shown for comparison in Figure \ref{fig:TP_profiles_comparison}.

We tested another retrieval case which included using a blackbody instead of the PHOENIX stellar photosphere model to compute \Fs{}. We find that in both alternative retrieval setups, we obtain consistent \PT{} profiles, sub-solar atmospheric metallicities and solar/sub-solar C/O ratios, and sub-solar abundances of \HtwoO{}. Furthermore, we also ran a free chemistry retrieval (\citep{madhusudhan_temperature_2009} profile, \HtwoO{} and CO abundance free) keeping planetary rotational velocity ($v_\mathrm{P}$ sin$i$) free. We find that even when keeping $v_\mathrm{P}$ sin$i$ free, our constraints are robust against potential degeneracy between the moelcular abundances and the broadening of spectral lines due to sources like atmospheric circulation or superroration. The results from these alternative models, shown in Appendix \ref{app:retrieval_madhu_seager}, \ref{app:Fs_Blackbody}, and \ref{app:retrieval_vsini_free}, strengthen the robustness of our constraints on the atmosphere of WASP-122b against a range of model assumptions.
We discuss the constraints on WASP-122b's thermal structure in the context of its sub-solar metallicity atmosphere in Section \ref{sec:discuss:met_TP}. 

%% Overplot of the TP profiles from different model assumptions

%%%%%%%%%%%%%%%%%%%%%%%%%%%%%%%%%%%%%%%%%%%%%%%%%%%%%%%%%%%
\subsection{\KpVsys{} Cross-Correlation Map from the Retrieved Model}
\label{sec:results:KpVsys_retrieved_model}
%%%%%%%%%%%%%%%%%%%%%%%%%%%%%%%%%%%%%%%%%%%%%%%%%%%%%%%%%%%
We measure \Kp{} = 204.91$_{-7.26}^{+7.37}$ \kms{} and \Vsys{} = 25.97$_{-4.22}^{+4.29}$ \kms{} from the equilibrium chemistry retrieval, which is consistent within 1$\sigma$ with the values constrained by the free chemistry retrieval (\Kp{} = 204.08$_{-4.92}^{+4.74}$ \kms{} and \Vsys{} = 26.34$_{-2.74}^{+2.92}$ \kms{}). The measured precision on the velocities is marginally lower than those from similar observations from IGRINS (e.g. \cite{smith_roasting_2024, brogi_roasting_2023} due to the lack of post-eclipse data. The best-fit values for both \Kp{} and \Vsys{} are shifted from the expected literature values (listed in Table \ref{tab:system_params}) by $\sim3\sigma$, which could be a signature of atmospheric dynamics and inhomogeneity (e.g. \cite{beltz_magnetic_2023, wardenier_phase-resolving_2024, wardenier_pre-transit_2025}). We discuss this in more detail in Section \ref{sec:discuss:KpVsys_shifts}.   

Following the same steps as in Section \ref{sec:cross_correlation_analysis}, but using the best fit retrieved model instead of the 1D-RCTE model, we compute the \KpVsys{} map for the CCF S/N and the log-likelihood derived confidence intervals. The resulting CCF S/N map shows a peak centred at the \Kp{} and \Vsys{} constrained by the retrieval, with the S/N of the CCF peak as 3$\sigma$, obtained by subtracting the median of the CCF \KpVsys{} map and dividing by its standard deviation. We note that the value of CCF S/N computed this way is not a very robust metric for the significance of detection, as it can be biased by the overall noise structure in the \KpVsys{} map. A better, more statistically motivated approach, is to instead use the log-likelihood values derived from the CCF themselves, as done for the RCTE model in Section \ref{sec:cross_correlation_analysis}. We find that the signal in log-likelihood space is tightly constrained, with the corresponding $\pm3\sigma$ confidence interval lying well within the $\pm3\sigma$ of the best-fit \Kp{} and \Vsys{} measured from the retrievals.  

We also checked for the contributions from individual molecular species in the model by following the same steps as \cite{smith_roasting_2024}, which involves computing the difference between the \KpVsys{} map corresponding to model with all the species and the \KpVsys{} map for the same model but a given species removed. We find that the \KpVsys{} map for \HtwoO{} shows a clear significant peak, but none of the other species in the model show a detection(as shown in Figure \ref{fig:kpvsys_all_species}). This aligns with the unconstrained posteriors obtained from the free-chemistry retrievals for all species except \HtwoO{}, indicating that the detection is driven primarily by \HtwoO{}. 

\section{Discussion}
\label{sec:discussion}

\subsection{Metallicity and thermal structure of WASP-122b}
\label{sec:discuss:met_TP}
%%% Overplot of the TP profiles with the contribution functions
\begin{figure*}
\centering
\includegraphics[width=1\textwidth]{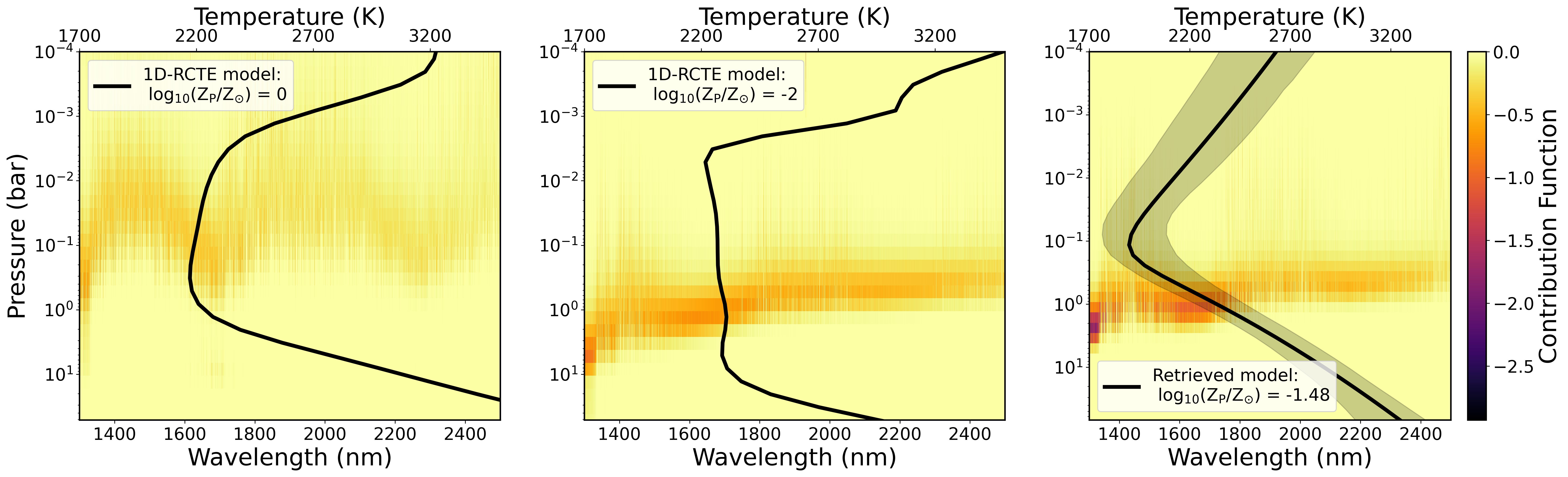}
\caption{Wavelength dependent contribution functions for the combination of \PT{} profile (overplotted) and metallicity for WASP-122b across three cases: (Left panel) 1D-RCTE model \PT{} profile with solar metallicity, (Middle panel) 1D-RCTE model \PT{} profile with $-$2 dex solar metallicity, and (Third panel) Retrieved \PT{} profile from the equilibrium chemistry retrieval, with the corresponding retrieved best fit metallicity of $-$1.48 dex (shaded region showing the $\pm$1$\sigma$ bounds obtained from the retrieved posteriors.). The darker regions in the map correspond to a higher contribution to the emission spectrum. Note that for the solar metallicity 1D-RCTE model, the contribution function spans a wide range of pressures from 1 to 0.001 bar, covering the inverted part of the \PT{} profile, whereas reducing the metallicity pushes the contribution function deeper in the atmosphere to a smaller range of pressures (10 to 1 bar), which for the case of the retrieved model leads to most of the spectral lines emanating from the deeper non-inverted part of the \PT{} profile causing them to be in absorption.}
\label{fig:TP_contribution_func}
\end{figure*}
Both free and equilibrium chemistry retrievals indicate that WASP-122b has an atmosphere with significantly sub-solar metallicity, and emission spectrum with lines in absorption rather than emission. We highlight that in the context of our retrieval constraints on the metallicity and \PT{} profile, the solar composition assumption for the 1D-RCTE model is incorrect. The solar composition 1D-RCTE model has all the spectral lines in emission in the IGRINS wavelength range. The wavelength dependent photosphere (the pressure at which the wavelength dependent optical depth $\tau(\lambda)$ = 2/3) of the 1D-RCTE model is centred roughly around 0.001 bar. In contrast, the retrieved best-fit model with best-fit sub-solar metallicity \logZpByZs{} = $-$1.48 dex has most of the spectral lines in absorption (as shown in in Figure \ref{fig:KpVsys_maps}), and its $\tau(\lambda)$ = 2/3 photosphere is centred deeper in the atmosphere around 0.1 bar. This helps to explain the observed anti-correlation in the CCF \KpVsys{} computed using the solar composition 1D-RCTE model, and a positive correlation with the best-fit retrieved model, as shown in Figure \ref{fig:KpVsys_maps}. Note that the pressures P($\lambda$)$_{\tau}=2/3$ only simply indicate pressure levels from which most of the photons are emitted at a given wavelength. In reality, the total flux emitted at a given wavelength comes from a range of pressure levels, and this can be visualized through a contribution function as defined by \cite{knutson_multiwavelength_2008}, adapted from Equations 4.4.1 to 4.4.8 in \cite{chamberlain_theory_1987} (for along the zenith, that is $\mu$ = cos$\theta$ = 1):

\begin{equation}
\label{eq:contribution_function}
cf(P,\lambda) = B(\lambda, T(P))\frac{\Delta e^{-\tau(\lambda, P)}}{\Delta \mathrm{log}(P)}
\end{equation}  

where $B(\lambda, T(P))$ is the blackbody emission expected at a given wavelength $\lambda$ and temperature $T(P)$ (at the given pressure level, defined by the \PT{} profile), and $\tau(\lambda, P)$ is the optical depth at the given pressure and wavelength. This form of contribution function essentially represents the intensity from a given pressure layer, assuming a Planck source function. The $\tau(\lambda, P)$ is governed by both \PT{} profile and the abundances of species with relatively high opacities in a given wavelength range. Following Equation \ref{eq:contribution_function}, we computed the contribution function for three cases: 1) 1D-RCTE \PT{} profile with \logZpByZs{} = 0 (solar composition), 2) 1D-RCTE \PT{} profile with \logZpByZs{} = $-$2 dex, and 3) Retrieved \PT{} profile with \logZpByZs{} = $-$1.48 dex. The contribution functions obtained for all three cases are shown in Figure \ref{fig:TP_contribution_func}. 

We find that for the solar composition 1D-RCTE case, the contribution to the emission spectrum comes from a wide range of pressures spanning 1 to 0.001 bar. In contrast, for both the 1D-RCTE model with \logZpByZs{} set to $-$2 dex, and $-$1.48 dex (best fit retrieved value), the contribution comes from a smaller range of pressures centred deeper in the atmosphere around 1 bar. This aligns with the intuitive expectation for an H/He dominated atmosphere like that of WASP-122b. Decreasing atmospheric metallicity leads to an average decrease in optical thickness across the wavelength range, which means that photons originating deeper in the atmosphere can still escape, with lesser chance of being absorbed as compared to the case of an atmosphere with higher optical thickness due to higher metallicity. The effect of decreasing metallicity on the self-consistently computed RCTE thermal structure for WASP-122b can be seen for the metallicity set to -2 dex in the middle panel in Figure \ref{fig:TP_contribution_func}. In contrast to the solar composition case which is largely inverted in $\sim$ 1 to 0.01 bar range, the \PT{} profile for the sub-solar case is non-inverted in the same range of pressures. This change can be explained by the fact that decreasing metallicity leads to an overall decrease in molecular (TiO, VO, FeH) and atomic species (e.g. Fe, Mg, and Na) which cause thermal inversion by absorbing the bulk of incoming optical stellar irradiation. Note however that decreasing the metallicity from solar to -2 dex solar is not enough to remove the inversion in the upper atmosphere (pressures $<$ 10$^{-3}$), where the inversion is driven by the onset of thermal dissociation of \HtwoO{}, which reduces the ability of the atmosphere to cool effectively by emitting radiation in the infrared.

\begin{figure}
\centering
\includegraphics[width=0.5\textwidth]{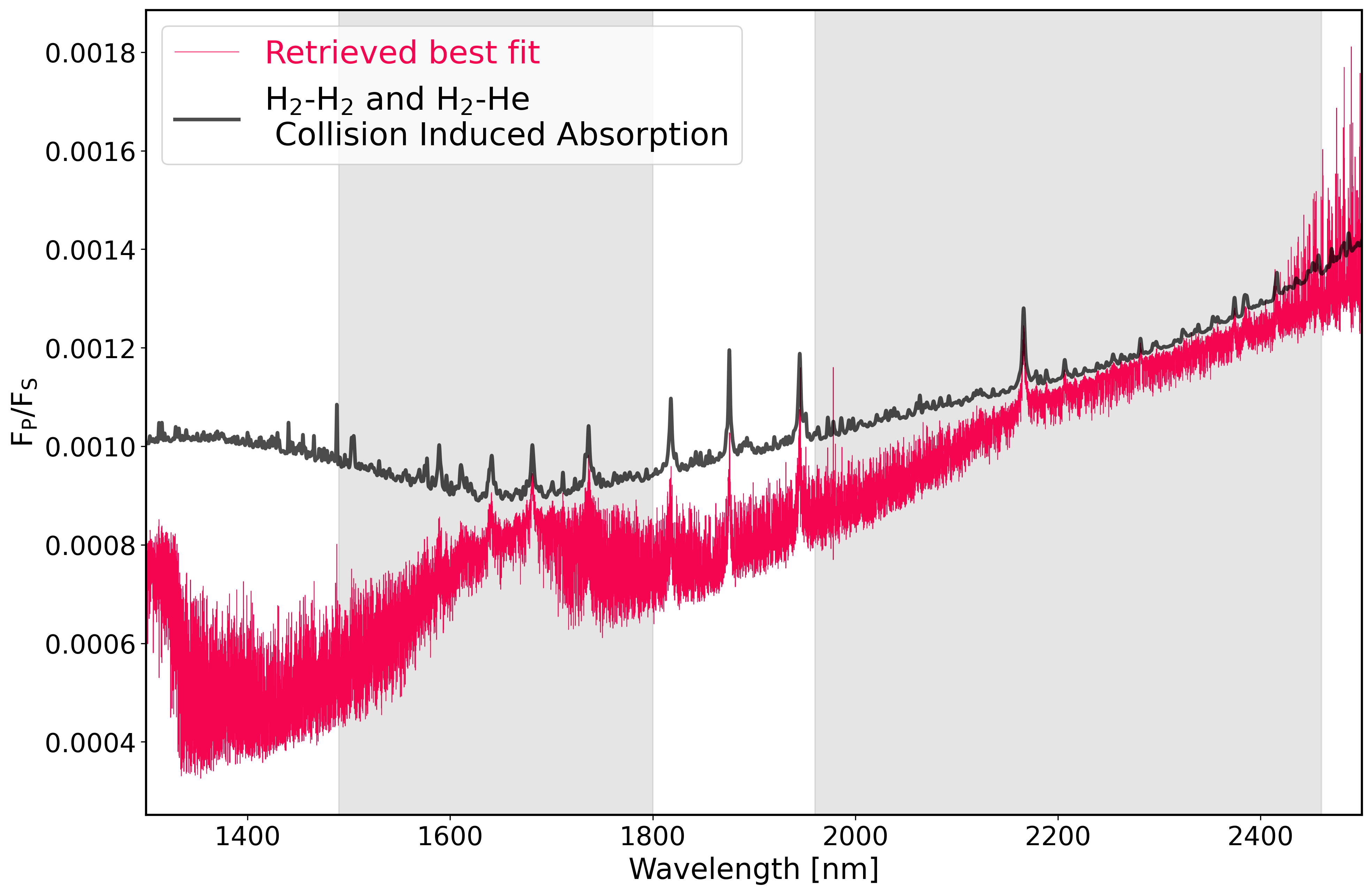}
\caption{Best fit emission spectrum model for WASP-122b obtained from the equilibrium chemistry retrieval in comparison with its corresponding H$_2$-H$_2$ and H$_2$-He collision induced absorption (CIA) contribution. The shaded region shows the H and K wavelength bands covered by the IGRINS observations in this work. The majority of the spectrum probed by our observations has absorption lines, except the reddest end of K band where there are a small number of lines in emission.}
\label{fig:best_fit_FpFs_CIA}
\end{figure}

As shown in Figure \ref{fig:TP_contribution_func}, a sub-solar metallicity atmosphere pushes the contribution function for the emission spectrum of WASP-122b deeper in the atmosphere (as compared to the RCTE solar metallicity model), around 1 bar, where the \PT{} profile is non-inverted. This is true for all the alternative \PT{} profile parametrizations we tested in Section \ref{sec:results:constraints}. This explains why the RCTE model led to an anticorrelated peak in the \KpVsys{} map, whereas the retrieved model yields a positive peak (as shown in Figure \ref{fig:KpVsys_maps}). The solar composition RCTE model has the contribution function spanning largely the inverted part of the\PT{} profile, leading to nearly all spectral lines in emission. In contrast, the retrieved best fit model with sub-solar metallicity has the contribution function restricted largely to the non-inverted part of the \PT{} profile, leading to the spectral lines emanating from the region around 1 bar to be in absorption. We note that since the contribution function for the best fit model is restricted to a narrow range in deep atmosphere, we obtain weaker constraint on the nature of the \PT{} profile in the upper atmosphere (as discussed previously in Section \ref{sec:results:constraints}). On inspecting the best fit emission spectrum model with its corresponding continuum set by the H$_2$-H$_2$ and H$_2$-He collision induced absorption (CIA) (Figure \ref{fig:best_fit_FpFs_CIA}, we indeed confirm that most of the spectral lines are in emission, and the weak constraint on the possible change to inversion in the \PT{} profile in the upper atmosphere (pressures $<$ 0.1 bar) comes from a very small number of lines in emission towards the reddest end of K band. 

In the context of the overall population of hot and ultra-hot Jupiters, the \PT{} profile constrained by our retrieval supports the idea that for equilibrium temperature similar to WASP-122b (T$_{\mathrm{eq}}$$\sim$1900 K), hot-Jupiters have possibly already begun transitioning from having non-inverted to inverted thermal structure in the upper atmosphere pressure ranges $<$0.1 bar. However, our retrievals only obtain the tightest constraint on the \PT{} profile in layers at pressure $>$0.1 bar, and stronger constraints on the upper atmosphere ($<$0.1 bar) are needed to confirm this transition. If confirmed, this would roughly agree with the results from \cite{baxter_evidence_2021} who analyzed a statistical sample of hot to ultra-hot Jupiter secondary eclipses observed using Spitzer 3.6 and 4.5 $\mu$m and find that the transition to inversions based on the emission from CO in the 4.5 $\mu$m channel happens around 1660 $\pm$ 100 K. Curiously, WASP-122b's metallicity of 0.033$_{-0.016}^{+0.018}$ $\times$ solar falls within the predicted range of metallicities (0.03 to 30 $\times$ solar) required to explain the scatter in the \HtwoO{} absorption features measured by HST/WFC3 for a sample of 19 hot and ultra-hot Jupiters \cite{mansfield_unique_2021}. 

Our findings for WASP-122b highlight the fact that whether an inverted or non-inverted \PT{} profile leads to spectral signatures in emission or absorption respectively depends on the atmospheric composition, which governs the range of pressures where the contribution function for the emission spectrum peaks. This is especially important for the high spectral resolution and wide wavelength range data, similar to the one presented in this work, which has adequate sensitivity to precisely constrain the line shapes and contrast. This was also indicated earlier by the high-resolution emission spectroscopy of HD 209458 b by \cite{schwarz_evidence_2015}, albeit in a narrower wavelength of CRIRES (2.305 to 2.33 $\mu$m) as compared to IGRINS (1.45 to 2.42 $\mu$m). \cite{schwarz_evidence_2015} show that a model corresponding to a strongly inverted \PT{} profile results in an anti-correlated peak in the CCF \KpVsys{} map similar to what we obtain for the 1D-RCTE profile, but a near-isothermal profile shows a positive peak, indicating the lack of strong thermal inversion on the day side of the planet. \cite{schwarz_evidence_2015} also note that a weakly inverted thermal profile yields spectral lines with their wings (probing the continuum) in absorption but the cores (probing lower pressures) in emission. Similar complex behaviour of mixed absorption and emission features from \HtwoO{} and \CO{} in the same spectrum (from 1.4 to 3.6 $\mu$m) due to the two species probing either inverted or non-inverted parts of the \PT{} profile was predicted by \cite{parmentier_thermal_2018} for WASP-121b, particularly relevant at high resolutions (see their Figure 8). We note, however, that the two recent works that have studied the atmosphere WASP-121b using IGRINS \citep{smith_roasting_2024} and CRIRES+ and ESPRESSO \citep{pelletier_crires_2025} measure a dayside emission spectrum with all lines including CO and \HtwoO{} CRIRES+ and ESPRESSO wavelength range are in emission, indicating that the dayside of the planet has an inverted thermal structure and solar to slightly super-solar composition atmosphere.

In summary, high-resolution spectroscopy data with wide wavelength coverage have enough information that a model with adequate flexibility in chemical abundance and \PT{} profile parametrizations can indeed capture the complexity of the underlying chemical composition and thermal structure. In this context, we also emphasize the importance of moving beyond a simple dichotomy of inversion /non-inversion leading to spectral emission/ absorption lines when considering the emission spectra of hot and ultra-hot Jupiters. The better approach is to instead use model parametrizations with enough flexibility for atmospheric composition and \PT{} profile, which together govern the nature of atmospheric spectral signatures in high-resolution cross-correlation spectroscopy data.

\subsection{Comparison with previous studies for WASP-122b}
\label{sec:discuss:compare_previous_studies}
\cite{stangret_obliquity_2024} conducted high-resolution spectroscopy of WASP-122b during its primary transit in the visible wavelength range using ESPRESSO, and found that the planet's orbit is aligned with the stellar spin axis (spin-orbit obliquity $\lambda$ = 0.09$^{+0.88}_{-0.90}$ deg). \cite{stangret_obliquity_2024} also conducted a search for a range of atomic and molecular species using cross-correlation, and found no evidence of absorption from any of the species, including \HtwoO{}. We compare our constraints on the dayside atmosphere of WASP-122b with the findings of \cite{stangret_obliquity_2024} for the terminator of the planet. A caveat with such comparison is that observations by \cite{stangret_obliquity_2024} cover visible band whereas our observations cover the near-infrared bands. Moreover, a direct one to one comparison of spectroscopy in transmission and emission is not straightforward because of the differences in the pressure and temperature ranges probed by both.  

The non-detection of refractory bearing species (e.g. Fe, Mg, Ti, TiO, VO) in the ESPRESSO visible band data at the planet's terminator is consistent with our non-detections of these species on the planet's dayside, both explained by the sub-solar metallicity of the planet. If these species are in trace amounts on their relatively hotter dayside of the planet, the cooler temperatures at the terminator are likely to cause them to condense and make them further undetectable. In the case of \HtwoO{}, the non-detection in ESPRESSO could potentially be explained by a combination of comparatively smaller number of spectral lines in the ESPRESSO wavelength range as compared to IGRINS and the sub-solar abundance of \HtwoO{} decreasing the strength of the lines. However, we emphasize that this is a qualitative explanation, and we encourage further exploration in the visible band with other instruments, in both transmission and emission, to test these hypotheses. 

\subsection{Implications for the formation history of WASP-122b}
\label{sec:discuss:planet_formation}
\begin{figure}
\centering
\includegraphics[width=0.5\textwidth]{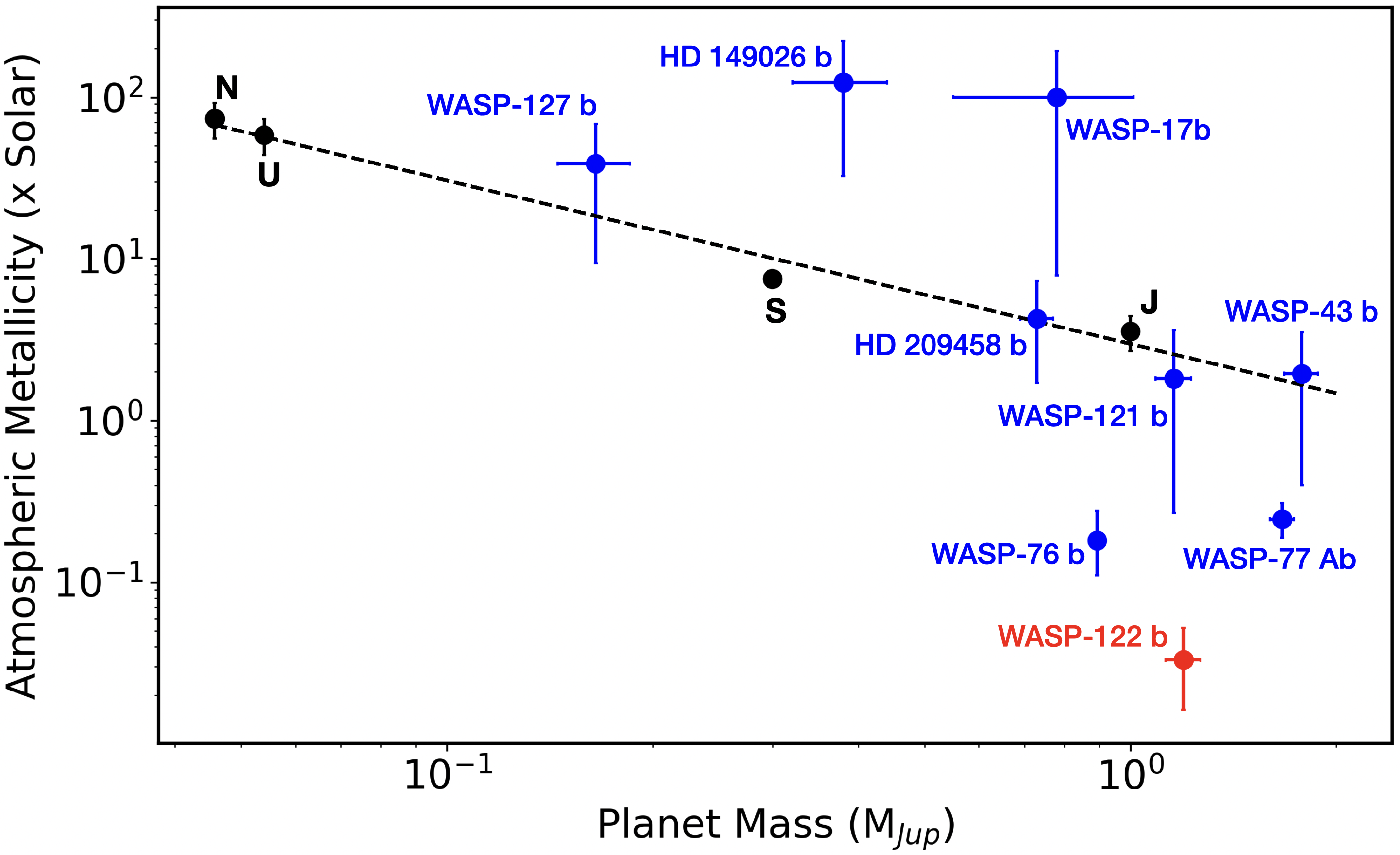}
\caption{Atmospheric metallicity of WASP-122b constrained from IGRINS observations in this work in the planet mass vs atmospheric metallicity space along with solar system gas-giants and other gas-giant exoplanets with well constrained metallicity measurements. The values for the solar system gas-giants have been taken from \protect\cite{guillot_giant_2022} (using C/H has proxy for metallicity). Sources for other planets from HRCCS and emission spectroscopy, HST/WFC3, and JWST are: via (C+O)/H: WASP-77Ab \protect\citep{smith_combined_2023}, WASP-127b \protect\citep{kanumalla_igrins_2024}, HD 149026b \protect\citep{bean_high_2023}, HD 209458b \citep{xue_jwst_2023},  WASP-76b \protect\citep{mansfield_metallicity_2024}; via O/H : WASP-17b \protect\cite{gressier_jwst-tst_2025}, WASP-43b \protect\cite{kreidberg_precise_2014, bartelt_measurement_2025}; via (C+O+R)/H, where R is total abundance of refractory elements: WASP-121b \protect\cite{smith_roasting_2024}. A linear slope fit to the solar system values is shown in dashed line. }
\label{fig:mass_metallicity}
\end{figure}

\begin{figure*}
\centering
\includegraphics[width=0.9\textwidth]{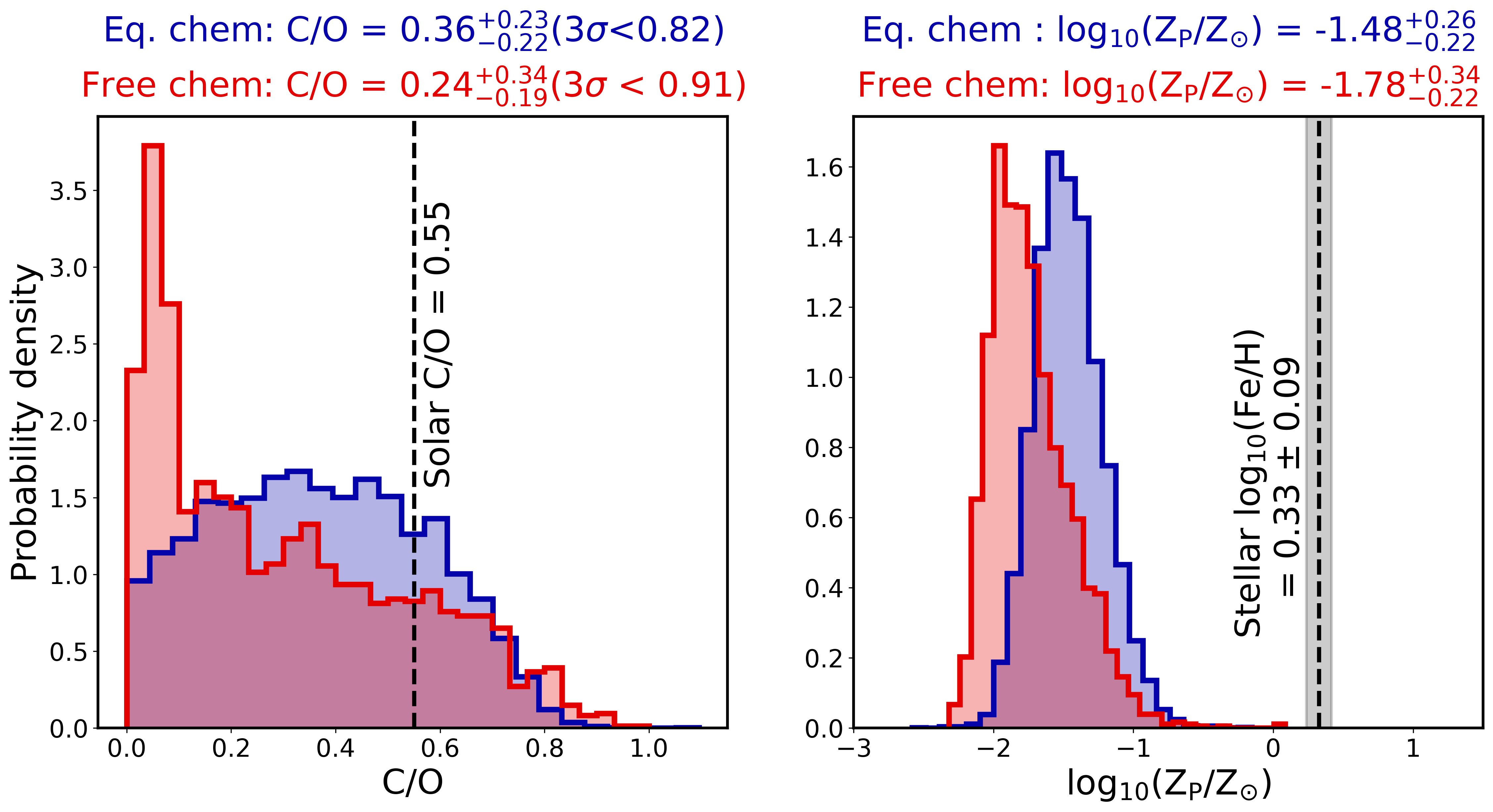}
\caption{1D posteriors for the C/O and metallicity of WASP-122b as compared to the solar and stellar values derived from the equilibrium chemistry retrieval (blue posteriors) and free chemistry retrieval (red posteriors).}
\label{fig:1D_C_to_O_and_metallicity}
\end{figure*}

Metallicity measurements for exoplanets through \HtwoO{} abundance measurements using HST \citep{kreidberg_precise_2014, welbanks_mass-metallicity_2019} and through \HtwoO{}, CO, and refractory measurements through ground-based high-resolution spectroscopy (e.g. \cite{line_solar_2021, smith_roasting_2024} indicate a departure from the decreasing metal enrichment with increasing planetary mass as observed in the solar system gas-giants (\cite{bean_high_2023, august_confirmation_2023}, also see Figure \ref{fig:mass_metallicity}). This suggests there is a diversity in possible formation pathways across the exoplanet population as compared to the solar system. Both equilibrium chemistry and free retrievals on the IGRINS data indicate that WASP-122b has a C/O ratio consistent with solar and metallicity \logZpByZs{} = $-$1.48$_{-0.22}^{+0.25}$ dex or 0.033$_{-0.016}^{+0.018}$ $\times$ solar. The metallicity of the host star measured by the discovery papers \cite{rodriguez_kelt-14b_2016} and \cite{turner_wasp-120_2016} (through Fe/H ratio from the visible band spectrum collected for radial velocity measurements) is log$_{10}$(Fe/H) = 0.326 $\pm$ 0.09 and log$_{10}$(Fe/H) = 0.32 $\pm$ 0.09 respectively, indicating that the host star is 2.09$\pm$0.43 $\times$ solar metal rich. This implies WASP-122b has 0.016$\pm$0.009 $\times$ the stellar metallicity, indicating that the planet's atmosphere is severely metal depleted relative to the host star. As shown in Figure \ref{fig:mass_metallicity}, the level of metal depletion observed for WASP-122b is lowest among all the other hot gas-giant exoplanets for which precise constraints on atmospheric metallicity have been obtained so far. 
%%% Mass metallicity diagram

WASP-122b is less massive as compared to WASP-77Ab (previously characterized using IGRINS \cite{line_solar_2021}), but shows an order of magnitude more metal depletion as compared to the latter, which has also been found to have an atmosphere with solar C/O ratio and sub-solar/stellar metallicity \citep{line_solar_2021, august_confirmation_2023, smith_combined_2023}. Similar to WASP-77Ab however, WASP-122b's metal depleted atmosphere with solar/sub-solar C/O ratio does not align with the broad predictions from standard core-accretion model of planet formation \cite{pollack_formation_1996}. A gas-giant that forms beyond the snow lines of C and O bearing molecules (\HtwoO{}, CO, \COtwo{}) and migrates after the disk dissipates is likely to end up with an elevated C/O ratio (following the C/O ratio of the gas in the disk) and low-metallicity atmosphere. In constrast, a gas-giant forming within the disk and migrating through it is likely to end up with O enriched atmosphere from accreting planetesimals, leading to low C/O and high-metallicity atmosphere \citep{oberg_effects_2011, madhusudhan_toward_2014, khorshid_simab_2022}. WASP-122b's solar/sub-solar C/O ratio and sub-solar metallicity, similar to that of WASP-77 Ab, doesn't align with either scenarios. Following the model predictions by \cite{khorshid_retrieving_2023} for WASP-77 Ab, it is possible that WASP-122b could have formed beyond the CO or \COtwo{} snow line in a disk with part of the carbon budget locked in the soot phase, and migrated to a location beyond the \HtwoO{} snow line before the disk dissipation.

Our upper limit on the C/O ratio to be solar/sub-solar is based on the non-detection of CO by both free and equilibrium chemistry retrievals. We note that the star WASP-122 is G4 type star with T$_\mathrm{eff}$ = 5802$\pm$95 K, which implies the stellar spectrum is expected to have significant CO absorption lines in the IGRINS wavelength range (especially in the K band) which could potentially contaminate the planetary CO signal. Our observations are taken during phases covering the planetary orbit well before the start of the secondary eclipse, which means that there should be an adequate shift in velocity space between the stellar and planetary CO lines causing negligible bias to the cross-correlation signal. We defer a more detailed simulation of the impact of stellar residuals on the retrievals to a future study.

\subsection{Interpreting the measured shifts in \Kp{} and \Vsys{}}
\label{sec:discuss:KpVsys_shifts}
Shifts in the peak velocity at which the planetary signal is measured as compared to the expected velocity from the known orbital parameters have been used as proxies for atmospheric dynamics in hot Jupiters \citep{beltz_significant_2021, savel_diagnosing_2023, brogi_roasting_2023, smith_roasting_2024, wardenier_phase-resolving_2024}. For both the best fit \Kp{} and \Vsys{} values from the retrievals (both free and equilibrium chemistry) we find that the measured planetary signal is shifted in velocity from the expected literature values. With respect to the expected \Kp{} and \Vsys{} values from \cite{rodriguez_kelt-14b_2016}, our equilibrium chemistry retrieval measures a positive $\Delta$\Kp{} = 17.1$\pm$8.1 \kms{}, and negative \Vsys{} with $\Delta$\Vsys{} =$-$8.75$\pm$4.5 \kms{}. Note that \Kp{} represents the rate of change of the Doppler shifts in the planet rest frame during the observation. The stronger the changes in Doppler shift in the planet rest frame, the larger the absolute value of $\Delta$\Kp{}. \cite{wardenier_pre-transit_2025} in their recently published GCM simulations find consistently negative $\Delta$\Kp{} driven by the planetary rotation. Negative $\Delta$Kp{} implies that the line-of-sight velocities become less blueshifted (or more redshifted) in the planet rest frame over the course of the observation due to tidally locked rotation, and a larger part of the dayside rotating into view during pre-eclipse. The positive $\Delta$\Kp{} measured by us, and its absolute value much larger than the equatorial rotation velocity of WASP-122b (4.5 \kms{}), is in tension with the model predictions from \cite{wardenier_pre-transit_2025}. The large magnitude of $\Delta$\Vsys{}, as compared to observed values of $\Delta$\Vsys{} e.g. $\sim$2 \kms{} for the planet WASP-76b which has a similar equatorial rotational velocity  \citep{wardenier_pre-transit_2025}, are also difficult to explain unless there are very strong winds, or an instrumental systematic effect.

It is worth noting here the relatively large uncertainties on both $\Delta$Kp{} and $\Delta$\Vsys{} derived solely from pre-eclipse data in this work. Since we only have pre-eclipse data for the planet, our posteriors for \Kp{} and \Vsys{} are highly correlated and hence the measured precision for both is low. Further observations covering post-eclipse phases are needed to improve the precision on both \Kp{} and \Vsys{} to confirm these findings, and shed further light on the differences from model predictions. 
% Augmenting the pre-eclipse data with post-eclipse data, especially using instruments with higher-resolving power like CRIRES+ could help in improving the precision on the measured phase-resolved Doppler shifts of the planetary signal and enable comparisons with predictions from atmospheric dynamics models.  
% \tbd{Any help interpreting these shifts is appreciated.}
% Since we only have pre-eclipse data for the planet, our posteriors for \Kp{} and \Vsys{} are highly correlated and hence the measured precision for both is low. Augmenting the pre-eclipse data with post-eclipse data, especially using instruments with higher-resolving power like CRIRES+ could help in improving the precision on the measured phase-resolved Doppler shifts of the planetary signal and enable comparisons with predictions from atmospheric dynamics models. 
\section{Conclusions}
\label{sec:conclusions}

In this work, we have presented high-resolution near-infrared emission spectroscopy observations of the hot-Jupiter WASP-122b, which occupies the transitional regime between hot to ultra-hot Jupiters. We used IGRINS on Gemini-South to observe the day-side of WASP-122b during pre-secondary eclipse orbital phases. Based on our analysis using established 1D HRCCS techniques, we conclude the following about the atmospheric chemical composition and thermal structure of the day-side of WASP-122b: 

\begin{itemize}
\item We find that the solar composition 1D-RCTE model for WASP-122b, which predicts an inverted \PT{} profile with all spectral lines in emission, yields an anti-correlated peak in the cross-correlation map close to the expected \Kp{} and \Vsys{} for the system. This indicates that the planetary signal in the data has lines largely in absorption rather than all in emission. 
\item Using a suite of retrievals with varying prescriptions for the chemistry (free and equilibrium), \PT{} profile parametrizations (B{\'e}zier spline and \cite{madhusudhan_temperature_2009}) and stellar spectrum (using PHOENIX and Blackbody), we find that WASP-122b has an atmosphere with significantly sub-solar metallicity \logZpByZs{} = $-$1.48$_{-0.22}^{+0.25}$ dex, solar/sub-solar C/O ratio (3$\sigma$ upper limit 0.82), and a non-inverted \PT{} profile. The measured sub-solar atmospheric metallicity is primarily driven by the cross-correlation signal from \HtwoO{}. We do not find significant contribution to the cross-correlation signal from any other molecular or atomic species. 
\item We find that the sub-solar metallicity of WASP-122b pushes the contribution function for the emission spectrum in the IGRINS wavelength range to a relatively smaller range of pressures, deeper in the atmosphere (as compared to the solar composition 1D-RCTE model) around 1 bar where the \PT{} profile is non-inverted. This leads to the \HtwoO{} spectral lines to be largely in absorption. Due to lack of contribution from upper layers of the atmosphere (pressures $<$ 10$^{-1}$ bars) for the retrieved best fit model, we do not obtain a strong constraint on the nature of thermal profile in that region.   
% \item The weakly inverted part of the \PT{} profile constrained by our retrieval for the upper atmosphere of WASP-122b (pressure $<$0.1 bar) supports the theoretical expectations that by equilibrium temperatures similar to WASP-122b ($\sim$1900 K), the irradiated gas-giant population has already begun transitioning from non-inverted to inverted thermal structures. However, there is not a clear dichotomy between inversion and non-inversion when it comes to the overall thermal structure. Emission or absorption features in the apparent day-side spectrum of hot-gas-giants will depend on the pressure ranges where the contribution function peaks. 
\item The sub-solar metallicity of WASP-122b constrained from the IGRINS observations is lowest among all gas-giant exoplanets with well constrained atmospheric metallicity so far. This combined with the solar/sub-solar C/O ratio makes WASP-122b inconsistent with the predictions from standard core-accretion model, and alternative scanrios are needed to explain its formation history. It is possible that WASP-122b formed beyond the CO/\COtwo{} snow line in a disk with part of the Carbon locked up in soot phase and migrated just beyond \HtwoO{} snow line before disk-dissipation. 
\item We find that the positive $\Delta$\Kp{} and the large magnitude of $\Delta$\Vsys{} constrained by us are inconsistent with the predictions from GCMs and typical expectation from planetary rotation respectively. However, due to the relatively large uncertainties on both measurements, further observations especially those covering post eclipse phases are needed to further investigate this inconsistency.   
\end{itemize}

In light of our main result in this work about interplay of atmospheric composition and \PT{} profile on the apparent spectra of hot-Jupiters, we highlight some final points relevant for WASP-122b, and in general for the future work on comparative studies of the hot and ultra-hot Jupiter population. Due to the limited orbital phase coverage of our observations (only pre-secondary eclipse), and relatively modest strength of the planetary signal in the data, we did not attempt to perform phase-resolved analyses to measure effects arising from spatially varying atmospheric composition and thermal structure from day to the night side of the planet. However, as recent works \citep{beltz_significant_2021, van_sluijs_carbon_2023, wardenier_phase-resolving_2024, wardenier_pre-transit_2025} have shown, the information content of HRCCS datasets are rich enough that with appropriate phase coverage and phase-resolved retrieval frameworks, these 3D variations in temperature, chemistry, and dynamics can indeed be measured through velocity shifts for individual chemical species. To probe these effects and obtain a more complete understanding of the role of 3D atmospheric properties on the apparent spectra, we recommend further observations of WASP-122b, covering the post-eclipse phases and transit in the near-infrared and visible band. These could also further provide insight into the positive $\Delta$\Kp{} and large $\Delta$\Vsys{} measured by us in this work. 

Finally, we emphasize that our results for WASP-122b in this work join the continuing efforts for detailed atmospheric characterization and comparative studies of hot and ultra-hot Jupiters. In this context, the high sensitivity to line-shapes and contrasts provided by the powerful combination of wide wavelength coverage (similar to IGRINS), high spectral resolution, and flexible retrieval model parametrizations, can play a crucial in measuring the complexities in temperature and compositional profiles and the various thermochemical mechanisms that affect them across the population of irradiated gas-giant exoplanets.  

\section*{Acknowledgements}
 We thank Luke Parker, Jayne Birkby, Anjali Piette, Spandan Dash, and Lennart van Sluijs for insightful discussions related to this work. Vatsal Panwar and Heather Cegla acknowledge support from the UKRI Future Leaders Fellowship grant (MR/S035214/1, MR/Y011759/1). Vatsal Panwar also acknowledges support from the UKRI Science and Technology Facilities Council (STFC) through the consolidated grant ST/X001121/1. M.R.L and K.K. acknowledge support from NSF grant AST-2307177. J.P.W. acknowledges support from the Canadian Space Agency (CSA) [24JWGO3A-03] and the Trottier Family Foundation via the Trottier Postdoctoral Fellowship held at the Institute for Research on Exoplanets (IREx). M.W.M. acknowledges support from the Heising-Simons Foundation through the 51 Pegasi b Fellowship program. P.C.B.S. acknowledges support provided by NASA through the NASA FINESST grant 80NSSC22K1598. This work is based on observations obtained at the international Gemini Observatory, a program of NSF’s NOIRLab, which is managed by the Association of Universities for Research in Astronomy (AURA) under a cooperative agreement with the National Science Foundation on behalf of the Gemini Observatory partnership: the National Science Foundation (United States), National Research Council (Canada), Agencia Nacional de Investigaci\'{o}n y Desarrollo (Chile), Ministerio de Ciencia, Tecnolog\'{i}a e Innovaci\'{o}n (Argentina), Minist\'{e}rio da Ci\^{e}ncia, Tecnologia, Inova\c{c}\~{o}es e Comunica\c{c}\~{o}es (Brazil), and Korea Astronomy and Space Science Institute (Republic of Korea). This work used the Immersion Grating Infrared Spectrometer (IGRINS) that was developed under a collaboration between the University of Texas at Austin and the Korea Astronomy and Space Science Institute (KASI) with the financial support of the US National Science Foundation 27 under grants AST-1229522 and AST-1702267, of the University of Texas at Austin, and of the Korean GMT Project of KASI. This work used open source software packages that include maptlotlib \citep{hunter_matplotlib_2007}, numpy \citep{harris_array_2020}, pymultinest \citep{buchner_x-ray_2014}, and scipy \citep{virtanen_scipy_2020}.

%%%%%%%%%%%%%%%%%%%%%%%%%%%%%%%%%%%%%%%%%%%%%%%%%%
\section*{Data Availability}
The raw science data included in this work are publicly available on the Gemini Observatory archive. The processed data, analysis outputs, and the code used to make the figures in the paper are available at \href{this URL}{https://doi.org/10.5281/zenodo.15847396}.


\bibliographystyle{mnras}
\bibliography{refs} % if your bibtex file is called example.bib

%\begin{thebibliography}{99}
%\end{thebibliography}

%%%%%%%%%%%%%%%%%%%%%%%%%%%%%%%%%%%%%%%%%%%%%%%%%%

%%%%%%%%%%%%%%%%% APPENDICES %%%%%%%%%%%%%%%%%%%%%

\appendix

\section{Full posteriors from the equilibrium chemistry and free chemistry retrievals}
\begin{figure*}
\centering
\includegraphics[width=1.05\textwidth]{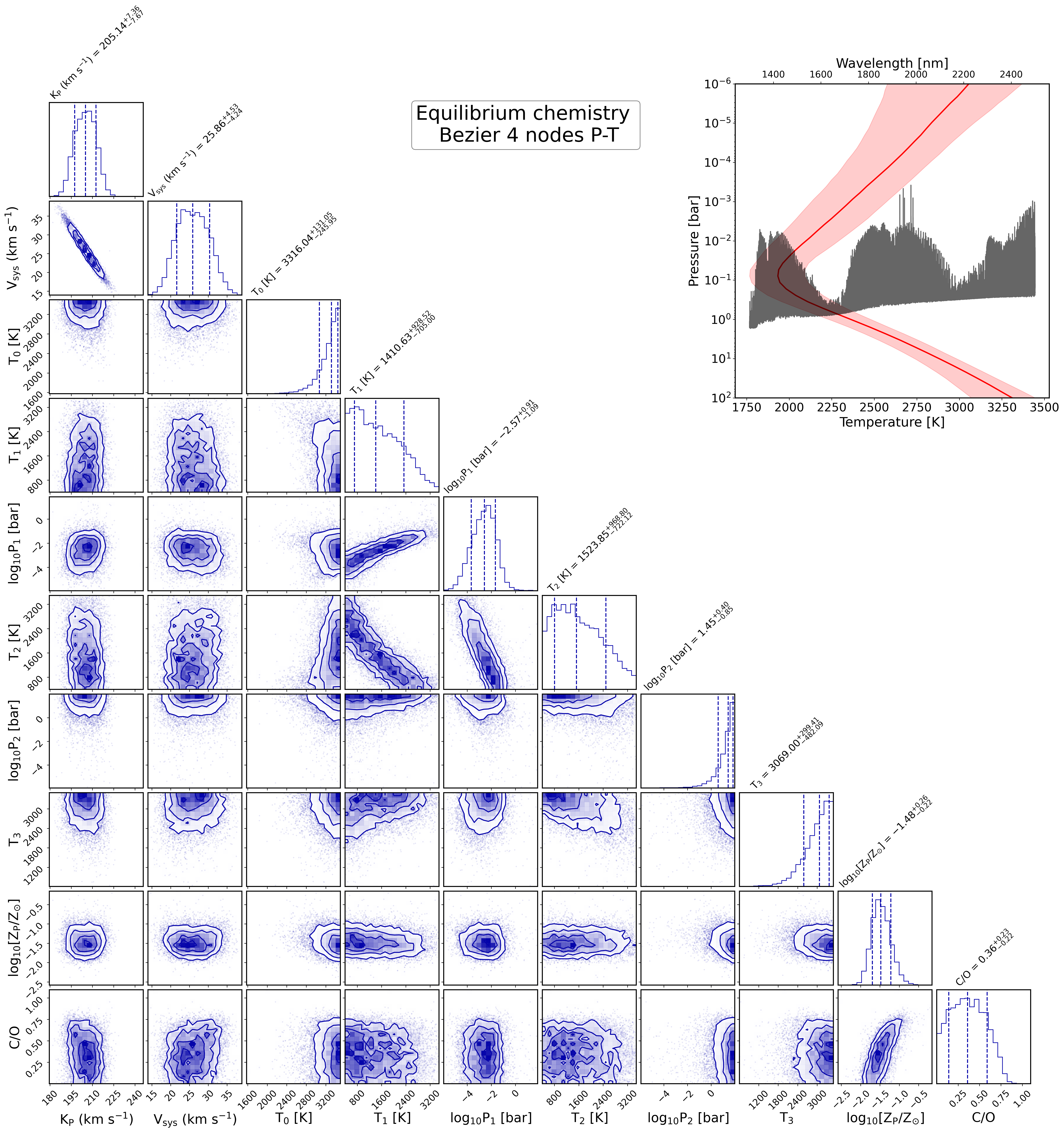}
\caption{Cornerplot showing the posterior distributions for all the free parameters in the equilibrium chemistry retrieval. The inset shows the best fit 4 node B{\'e}zier spline \PT{} profile, with the $\pm$1$\sigma$ bounds obtained from the posteriors, with the $\tau$=2/3 pressure levels for each wavelength overplotted in grey.}
\label{fig:full_eq_chem_retrieval_posteriors}
\end{figure*}

\begin{figure*}
\centering
\includegraphics[width=1.05\textwidth]{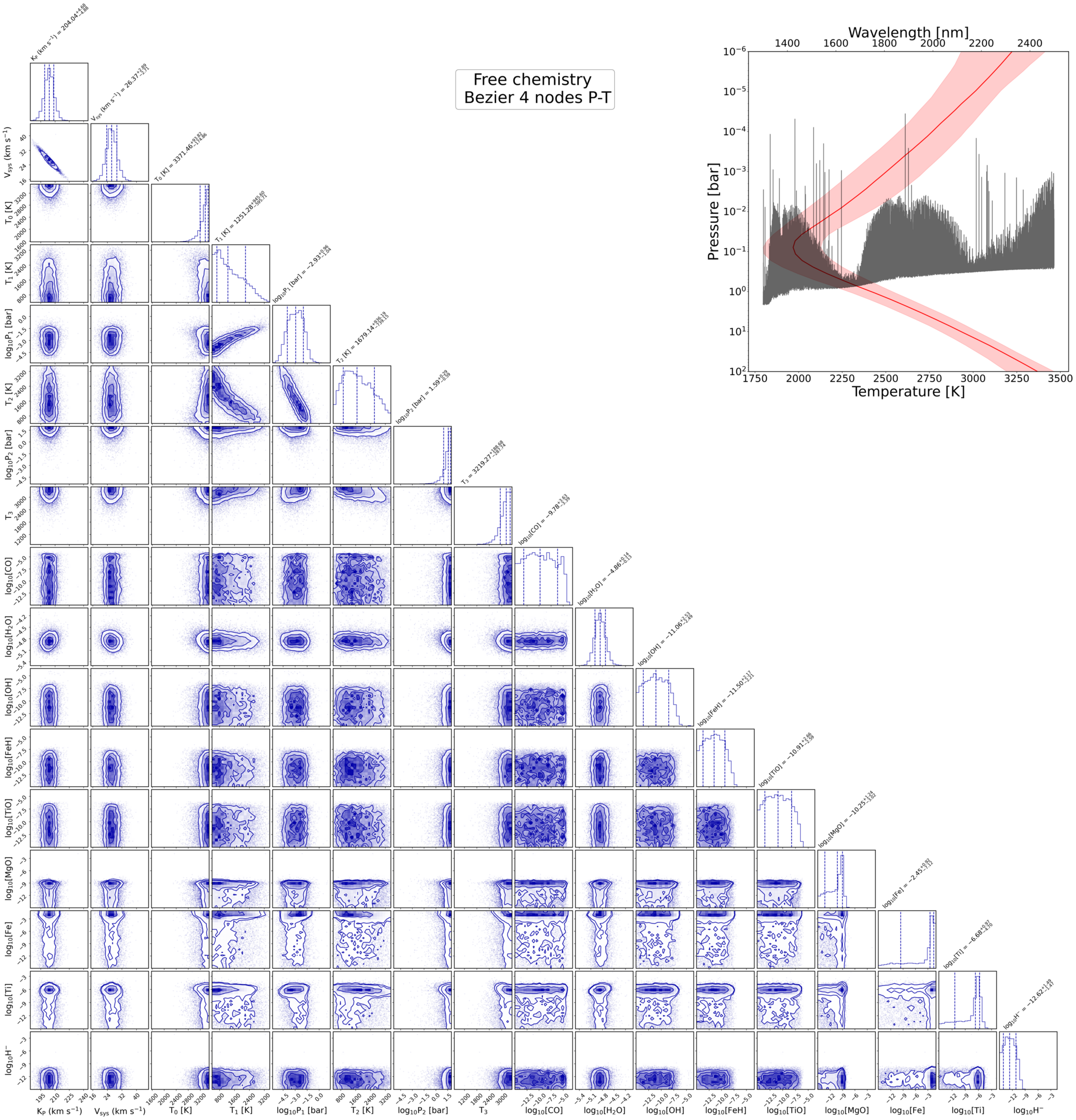}
\caption{Same as Figure \ref{fig:full_eq_chem_retrieval_posteriors} but for the free chemistry retrieval.}
\label{fig:full_free_chem_retrieval_posteriors}
\end{figure*}

In this section, we show the full posterior distributions obtained from our retrievals assuming free chemistry and equilibrium (assuming 4 nodes B{\'e}zier spline \PT{} profile) as described in Section \ref{sec:retrieval_analysis}. Figure \ref{fig:full_eq_chem_retrieval_posteriors} and \ref{fig:full_free_chem_retrieval_posteriors} show the cornerplots for all the free parameters in the equilibrium and free chemistry retrievals respectively. 

\label{app:full_posteriors}

\section{Testing the effect of alternative \PT{} profile parametrizations}
\label{app:retrieval_madhu_seager}
\begin{figure*}
\centering
\includegraphics[width=1.05\textwidth]{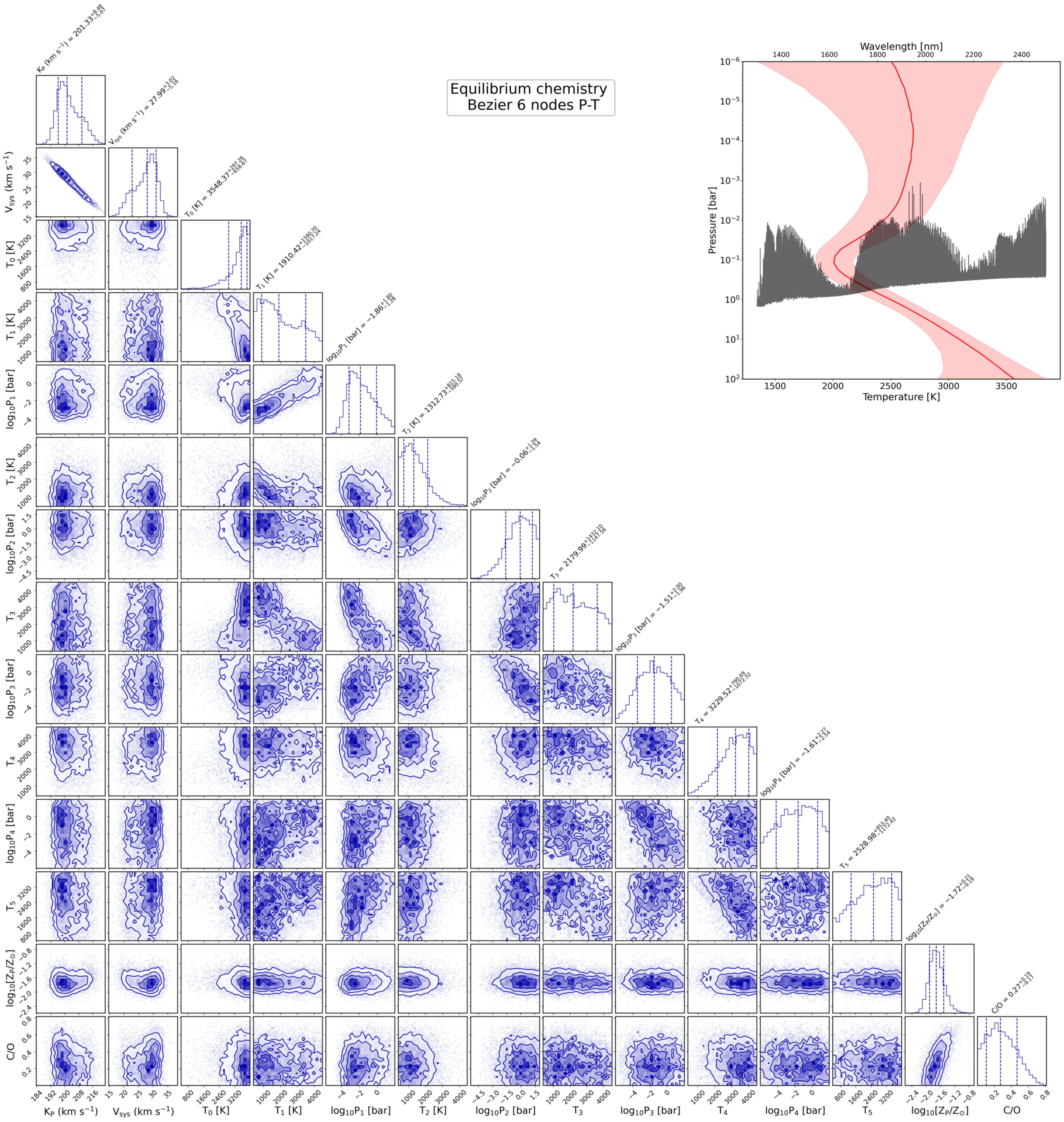}
\caption{Same as Figure \ref{fig:full_eq_chem_retrieval_posteriors}, but when using a 6 node B{\'e}zier spline \PT{} profile.}
\label{fig:full_eq_chem_Bezier_6}
\end{figure*}

\begin{figure*}
\centering
\includegraphics[width=1.05\textwidth]{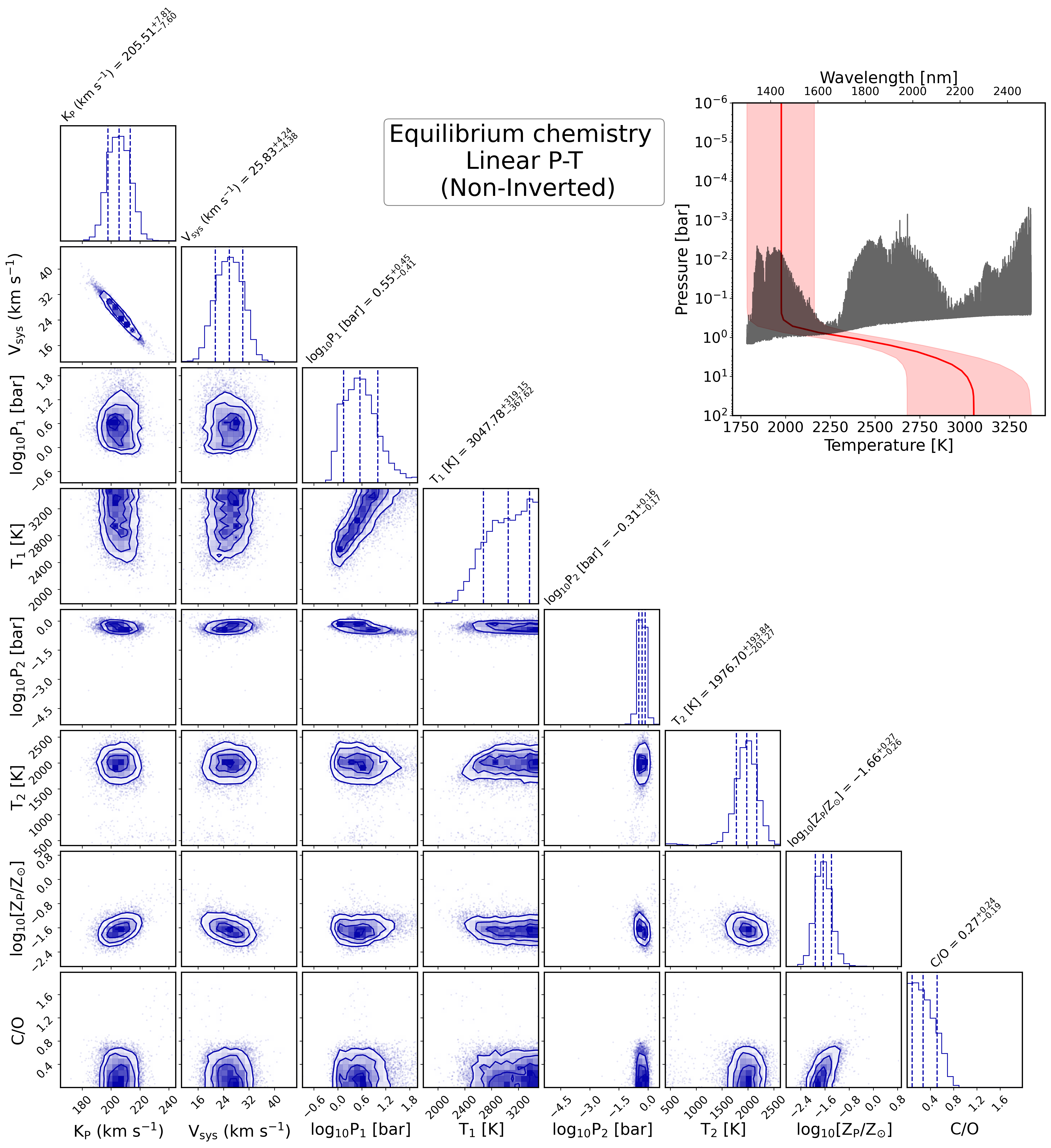}
\caption{Same as Figure \ref{fig:full_eq_chem_retrieval_posteriors}, but when using a linear \PT{} parametrization with no inversion.}
\label{fig:full_eq_chem_retrieval_posteriors_Linear_NON_INVERTED}
\end{figure*}

\begin{figure*}
\centering
\includegraphics[width=1.05\textwidth]{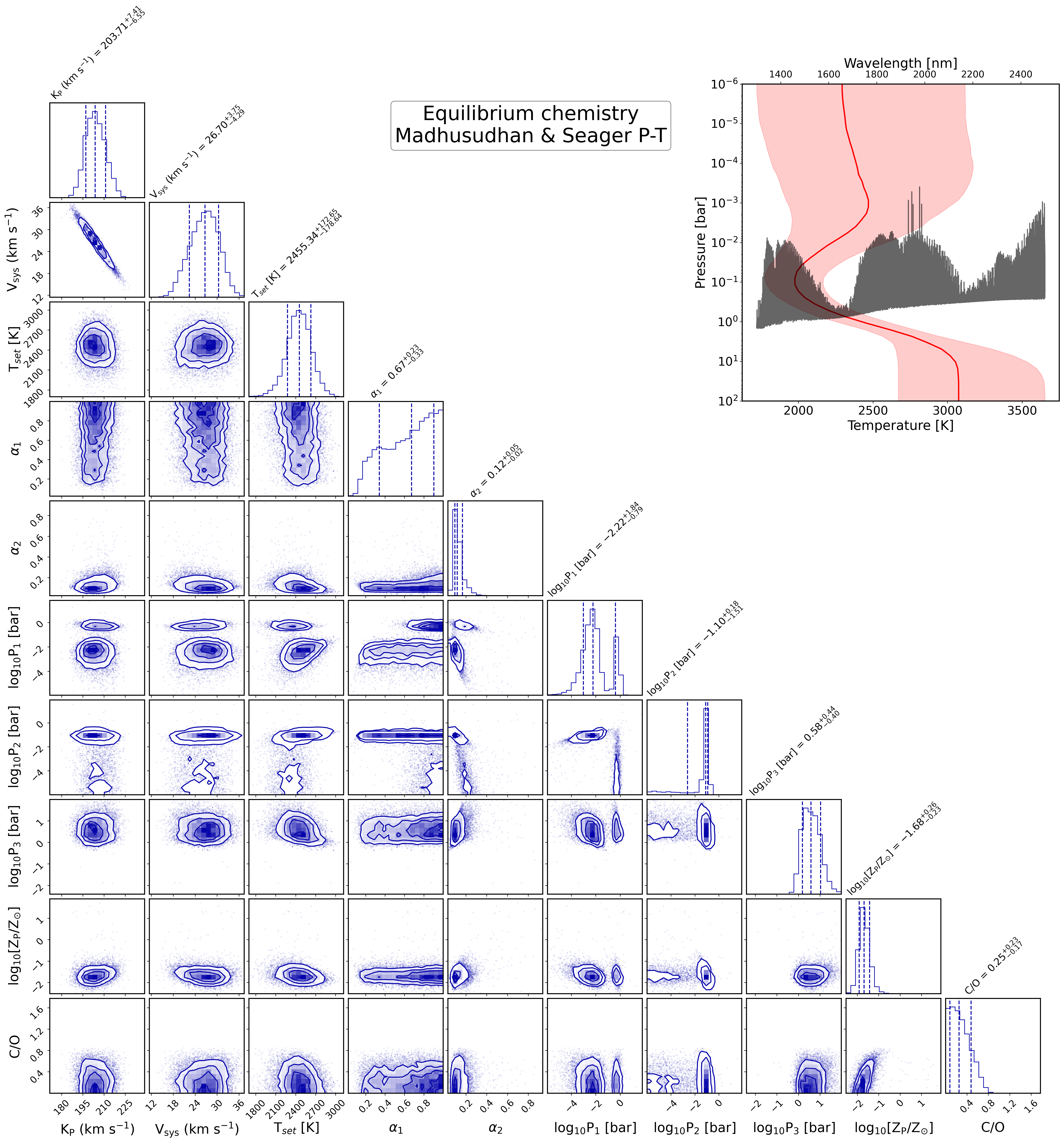}
\caption{Same as Figure \ref{fig:full_eq_chem_retrieval_posteriors}, but when using the \protect\cite{madhusudhan_temperature_2009} parametrization for the \PT{} profile.}
\label{fig:full_eq_chem_retrieval_posteriors_MadhuSeager}
\end{figure*}

To test the robustness of our retrieval results against the choice of \PT{} profile parametrization, we performed an equilibrium chemistry retrieval with identical setup as described in Section \ref{sec:retrieval_analysis}, but with three alternative \PT{} profile parametrizations: 1) 6 nodes B{\'e}zier spline profile, 2) Linear \PT{} with no inversion, and 3) the 6 parameter prescription introduced by \cite{madhusudhan_temperature_2009}. The full posteriors from 1) are shown in Figure \ref{fig:full_eq_chem_Bezier_6}, from 2) in Figure \ref{fig:full_eq_chem_retrieval_posteriors_Linear_NON_INVERTED}, and from 3) in \ref{fig:full_eq_chem_retrieval_posteriors_MadhuSeager}. The best fit parameters and their uncertainties are shown in Tables \ref{tab:eq_params_Bezier_6}, \ref{tab:eq_params_fit_Linear_TP}, and \ref{tab:eq_params_fit_madhu_seager} respectively.

In summary, the best fit parameters for the \Kp{}, \Vsys{}, C/O ratio, and metallicity (\logZpByZs{}) from all the alternative parametrizations are consistent with the equilibrium chemistry retrieval which uses the 4 node B{\'e}zier spline \PT{} parametrization. All the profiles follow the same non-inverted gradient in the same pressure ranges (0.1 to 1 bar) as the 4 node B{\'e}zier profile. However, in the upper atmosphere ($<$0.1 bar), all three alternative profiles are nearly isothermal and show a weaker inversion than the 4 node B{\'e}zier profile. This indicates that the contraint from the data on the thermal structure, regardless of the \PT{} parametrization, are the tightest in the high pressures ranging from 0.1 to 1 bar, This is consistent with the sub-solar metallicity consistently obtained from all the \PT{} parametrizations tested.

%%%%%%%%%% Equilibrium chemistry with 6 node Bezier
\begin{table}
\centering
\renewcommand{\arraystretch}{1.2} %
\caption{Best fit and 1$\sigma$ values for the parameters from equilibrium chemistry retrieval using the 6 node B{\'e}zier spline \PT{} profile parametrization.}
\begin{tabular}{|c|c|c|}
\hline
K$_{\mathrm{P}}$ (km s$^{-1}$) & [160.00, 250.00] & 201.31$_{-5.10}^{+8.45}$ \\
V$_{\mathrm{sys}}$ (km s$^{-1}$) & [5.00, 50.00] & 28.00$_{-5.14}^{+3.02}$ \\
T$_{0}$ [K] & [400.00, 4500.00] & 3548.80$_{-658.11}^{+296.66}$ \\
T$_{1}$ [K] & [400.00, 4500.00] & 1923.14$_{-1032.37}^{+1576.35}$ \\
log$_{10}$P$_{1}$ [bar] & [-6.00, 2.00] & -1.85$_{-1.28}^{+1.81}$ \\
T$_{2}$ [K] & [400.00, 4500.00] & 1314.07$_{-605.41}^{+814.67}$ \\
log$_{10}$P$_{2}$ [bar] & [-6.00, 2.00] & -0.08$_{-1.55}^{+1.32}$ \\
T$_{3}$ [K] & [400.00, 4500.00] & 2177.19$_{-1155.19}^{+1447.29}$ \\
log$_{10}$P$_{3}$ [bar] & [-6.00, 2.00] & -1.49$_{-1.98}^{+2.01}$ \\
T$_{4}$ [K] & [400.00, 4500.00] & 3232.03$_{-1075.56}^{+791.52}$ \\
log$_{10}$P$_{4}$ [bar] & [-6.00, 2.00] & -1.62$_{-2.54}^{+2.25}$ \\
T$_{5}$ [K] & [400.00, 4500.00] & 2530.93$_{-1175.21}^{+957.25}$ \\
log$_{10}$[Z$_{\mathrm{P}}$/Z$_{\odot}$] & [-6.00, 2.00] & -1.72$_{-0.19}^{+0.21}$ \\
C/O & [0.00, 2.00] & 0.27$_{-0.17}^{+0.20}$ \\
\hline
\end{tabular}
\label{tab:eq_params_Bezier_6}
\end{table}

%%%%%%%%%% Equilibrium chemistry with the Madhusudhan-Seager parametrization for the TP profile 
\begin{table}
\centering
\renewcommand{\arraystretch}{1.2} %
\caption{Best fit and 1$\sigma$ values for the parameters from equilibrium chemistry retrieval using the \protect\cite{madhusudhan_temperature_2009} \PT{} profile parametrization.}
\begin{tabular}{|c|c|c|}
\hline
Parameter & Prior Bounds & Best-fit $\pm$ 1$\sigma$ \\
\hline
K$_{\mathrm{P}}$ (km s$^{-1}$) & [160.00, 250.00] & 203.71$_{-6.56}^{+7.42}$ \\
V$_{\mathrm{sys}}$ (km s$^{-1}$) & [10.00, 50.00] & 26.69$_{-4.28}^{+3.76}$ \\
T$_{\mathrm{set}}$ [K] & [1000.00, 3500.00] & 2455.71$_{-179.08}^{+174.17}$ \\
$\alpha_{1}$ & [0.02, 1.00] & 0.68$_{-0.33}^{+0.23}$ \\
$\alpha_{2}$ & [0.02, 1.00] & 0.12$_{-0.02}^{+0.05}$ \\
log$_{10}$P$_{1}$ [bar] & [-6.00, 2.00] & -2.22$_{-0.78}^{+1.84}$ \\
log$_{10}$P$_{2}$ [bar] & [-6.00, 2.00] & -1.10$_{-1.57}^{+0.18}$ \\
log$_{10}$P$_{3}$ [bar] & [-6.00, 2.00] & 0.58$_{-0.40}^{+0.44}$ \\
log$_{10}$[Z$_{\mathrm{P}}$/Z$_{\odot}$] & [-6.00, 2.00] & -1.68$_{-0.23}^{+0.26}$ \\
C/O & [0.00, 2.00] & 0.25$_{-0.17}^{+0.23}$ \\
\hline
\end{tabular}
\label{tab:eq_params_fit_madhu_seager}
\end{table}

\section{Testing the effect of computing \Fs{} assuming blackbody radiation instead of PHOENIX stellar model}
\label{app:Fs_Blackbody}
\begin{figure*}
\centering
\includegraphics[width=1.05\textwidth]{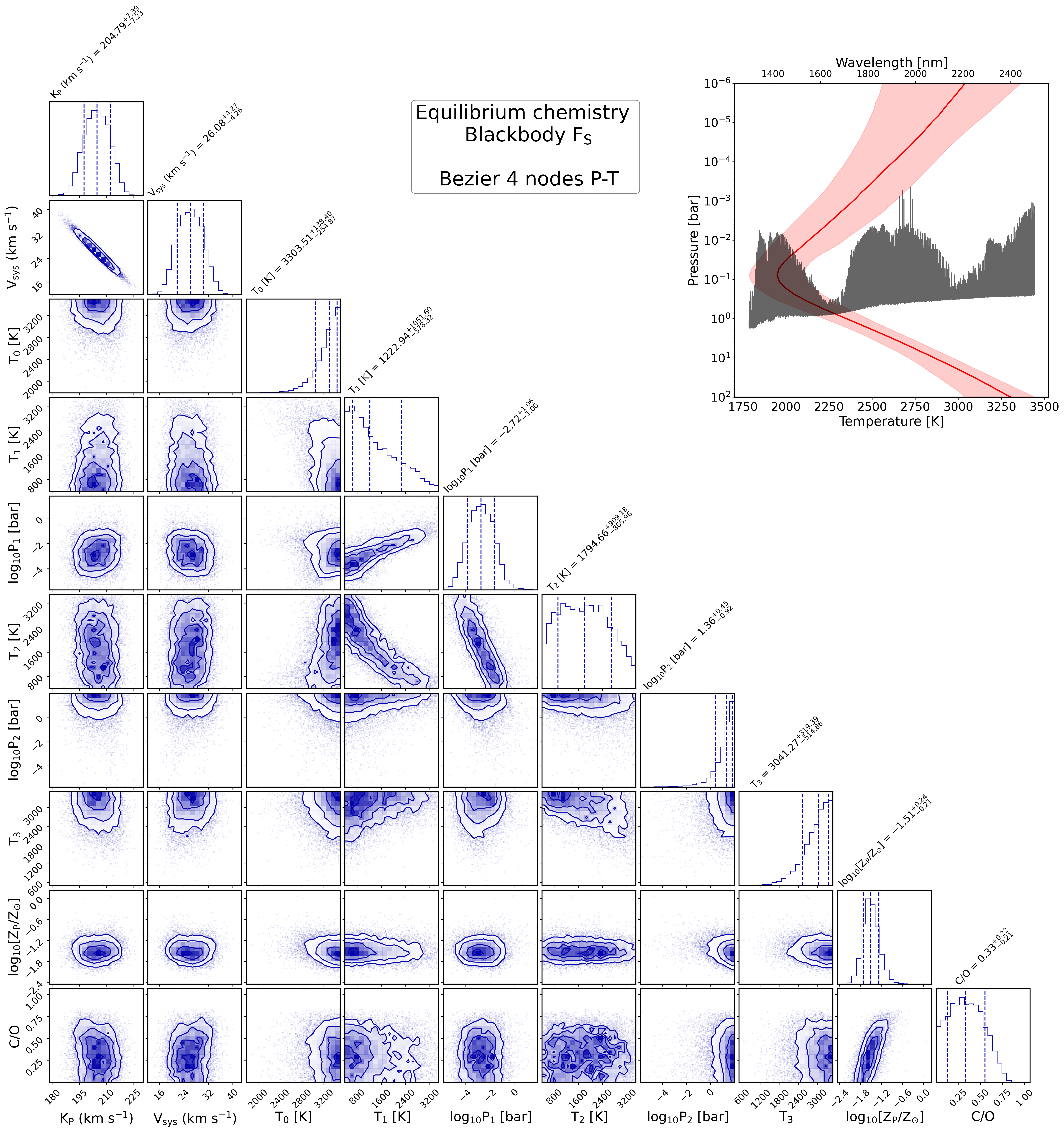}
\caption{Same as Figure \ref{fig:full_eq_chem_retrieval_posteriors}, but when using Blackbody corresponding to the T$_{\mathrm{eff}}$ of the host star instead of PHOENIX model for \Fs{} when computing \FpFs{}. 
}
\label{fig:full_eq_chem_retrieval_posteriors_Blackbody}
\end{figure*}

Since the host star WASP-122 is a G4 spectral type, the stellar spectrum \Fs{} is expected to have CO lines with non-negligible depths, which means the choice of how we consider for \Fs{} when computing \FpFs{} could potentially bias the retrieval when searching for CO in the planetary spectrum. To test our robustness against the choice of \Fs{}, we repeated the equilibrium chemistry retrieval with the same setup as described in Section \ref{sec:retrieval_analysis}, but with \Fs{} calculated as a blackbody spectrum corresponding to the stellar T$_{\mathrm{eff}}$ = 5802 K. The posteriors from this retrieval and the best fit parameters with their $\pm1\sigma$ uncertainties are shown in Figure \ref{fig:full_eq_chem_retrieval_posteriors_Blackbody} and Table \ref{tab:eq_params_fit_blackbody} respectively. In summary, the choice of whether we compute \Fs{} using a PHOENIX model or assuming blackbody emissiom, does not affect our retrieved constraints on the C/O ratio, metallicity, and the \PT{} profile.

%%%%%%%%%% Equilibrium chemistry with using Blackbody for the stellar model.
\begin{table}
\centering
\renewcommand{\arraystretch}{1.2} %
\caption{Best fit and 1$\sigma$ values for the parameters from equilibrium chemistry retrieval assuming \Fs{} to be a blackbody spectrum corresponding to the T$_{eff}$ of the star.}
\begin{tabular}{|c|c|c|}
\hline
Parameter & Prior Bounds & Best-fit $\pm$ 1$\sigma$ \\
\hline
K$_{\mathrm{P}}$ (km s$^{-1}$) & [160.00, 250.00] & 204.77$_{-7.27}^{+7.45}$ \\
V$_{\mathrm{sys}}$ (km s$^{-1}$) & [5.00, 50.00] & 26.08$_{-4.28}^{+4.30}$ \\
T$_{0}$ [K] & [400.00, 3500.00] & 3303.77$_{-255.54}^{+138.91}$ \\
T$_{1}$ [K] & [400.00, 3500.00] & 1215.79$_{-571.63}^{+1062.63}$ \\
log$_{10}$P$_{1}$ [bar] & [-6.00, 2.00] & -2.72$_{-1.07}^{+1.07}$ \\
T$_{2}$ [K] & [400.00, 3500.00] & 1791.86$_{-870.88}^{+926.09}$ \\
log$_{10}$P$_{2}$ [bar] & [-6.00, 2.00] & 1.37$_{-0.94}^{+0.45}$ \\
T$_{3}$ [K] & [400.00, 3500.00] & 3042.74$_{-523.37}^{+320.04}$ \\
log$_{10}$[Z$_{\mathrm{P}}$/Z$_{\odot}$] & [-6.00, 2.00] & -1.51$_{-0.21}^{+0.24}$ \\
C/O & [0.00, 2.00] & 0.34$_{-0.21}^{+0.22}$ \\
\hline
\end{tabular}
\label{tab:eq_params_fit_blackbody}
\end{table}

\section{Testing the effect of fitting for planetary rotational velocity}
\label{app:retrieval_vsini_free}
\begin{figure*}
\centering
\includegraphics[width=1.05\textwidth]{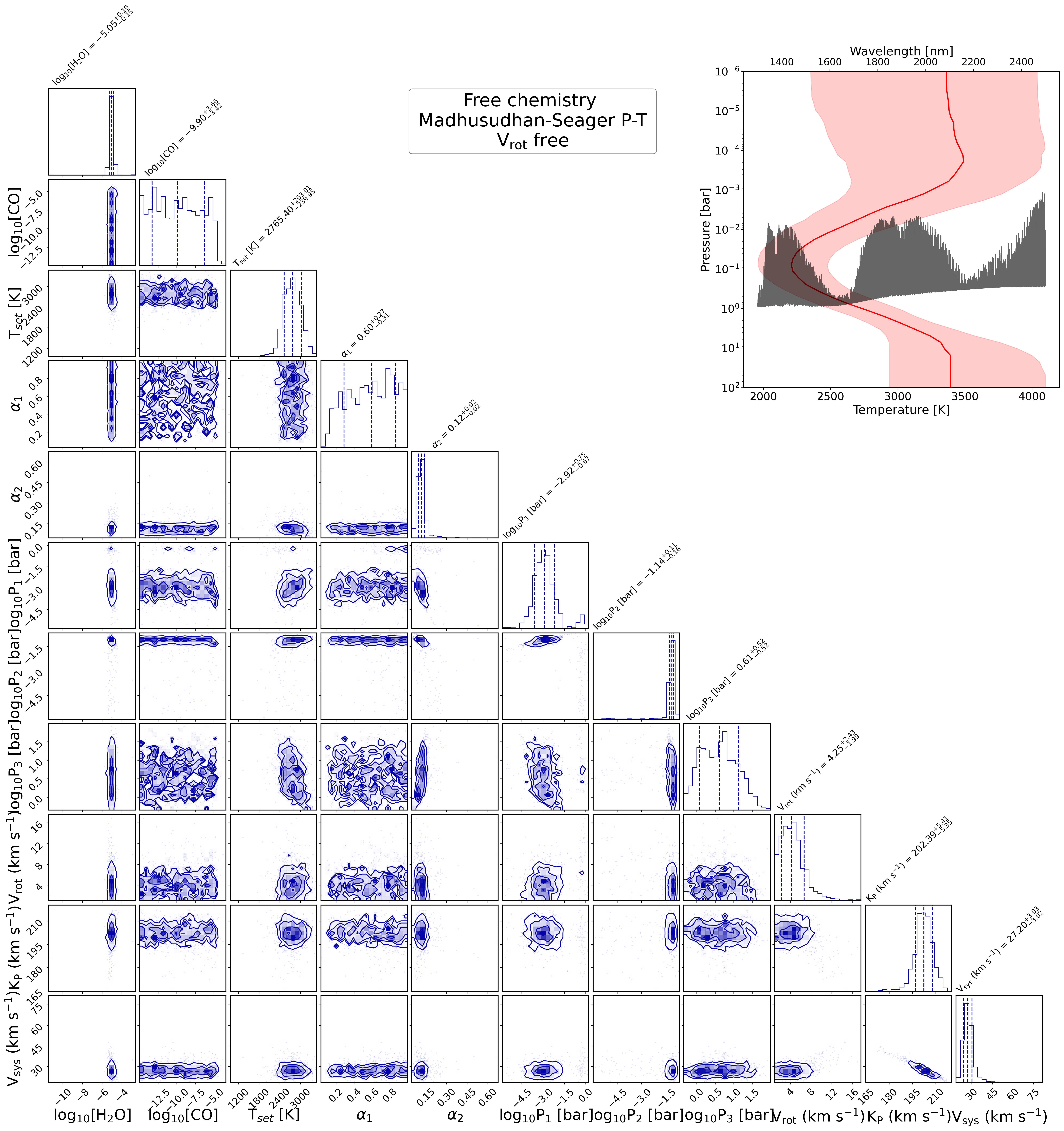}
\caption{Same as Figure \ref{fig:full_eq_chem_retrieval_posteriors_MadhuSeager}, but assuming free chemistry and also fitting for the planetary rotational velocity (v$_{\mathrm{rot}}$ = $v_\mathrm{P}$ sin$i$.}
\label{fig:free_chem_vsini}
\end{figure*}

The precise estimation of absolute abundance of species like \HtwoO{} depends heavily on the ability of the retrievals to fit the exact line profile. However, it is possible that there is a degeneracy between the abundance of water and the broadening of lines due to other sources like atmospheric circulation or superroration, and that the inferred low abundance of \HtwoO{} might be a compensating factor for extra broadening of lines from these effects. To test this, we ran a free chemistry retrieval with \HtwoO{} and CO abundance as free parameters, using \citep{madhusudhan_temperature_2009} profile, and also keeping $v_\mathrm{P}$ sin$i$ of the planet (which broadens the spectral lines) as a free parameter. We find that the constrained abundance of water remains unchanged even when allowing $v_\mathrm{P}$ sin$i$ to be free. The posteriors from this retrieval are shown in Figure \ref{fig:free_chem_vsini}, and the best fit values for each parameter and their $\pm$1$\sigma$ uncertainties are shown in Table \ref{tab:free_chem_vsini}.

%%%%%%%%%% Equilibrium chemistry with using Blackbody for the stellar model.
\begin{table}
\centering
\renewcommand{\arraystretch}{1.2} %
\caption{Best fit and 1$\sigma$ values for the parameters from free chemistry retrieval with the \protect\cite{madhusudhan_temperature_2009} \PT{} profile, and also fitting for planetary rotational velocity (v$_{\mathrm{rot}}$ = $v_\mathrm{P}$ sin$i$.)}
\begin{tabular}{|c|c|c|}
\hline
Parameter & Prior Bounds & Best-fit $\pm$ 1$\sigma$ \\
\hline
K$_{\mathrm{P}}$ (km s$^{-1}$) & [160.00, 250.00] & 202.39$_{-5.35}^{+5.41}$ \\
V$_{\mathrm{sys}}$ (km s$^{-1}$) & [10.00, 50.00] & 27.20$_{-3.02}^{+3.03}$ \\
T$_{\mathrm{set}}$ [K] & [1000.00, 3500.00] & 2765.40$_{-239.95}^{+263.01}$ \\
$\alpha_{1}$ & [0.02, 1.00] & 0.60$_{-0.31}^{+0.27}$ \\
$\alpha_{2}$ & [0.02, 1.00] & 0.12$_{-0.02}^{+0.02}$ \\
log$_{10}$P$_{1}$ [bar] & [-6.00, 2.00] & -2.92$_{-0.67}^{+0.75}$ \\
log$_{10}$P$_{2}$ [bar] & [-6.00, 2.00] & -1.14$_{-0.16}^{+0.11}$ \\
log$_{10}$P$_{3}$ [bar] & [-6.00, 2.00] & 0.61$_{-0.52}^{+0.52}$ \\
log$_{10}$[CO] & [-15.00, -1.00] & -9.90$_{-3.42}^{+3.66}$ \\
log$_{10}$[H$_{2}$O] & [-15.00, -1.00] & -5.05$_{-0.15}^{+0.19}$ \\
v$_{\mathrm{rot}}$ (km s$^{-1}$) & [1.00, 50.00] & 4.25$_{-1.99}^{+2.43}$ \\
\hline
\end{tabular}
\label{tab:free_chem_vsini}
\end{table}

\section{\KpVsys{} confidence interval maps for individual species}
\label{app:kpvsys_all_species}
\begin{figure*}
\centering
\includegraphics[width=1.05\textwidth]{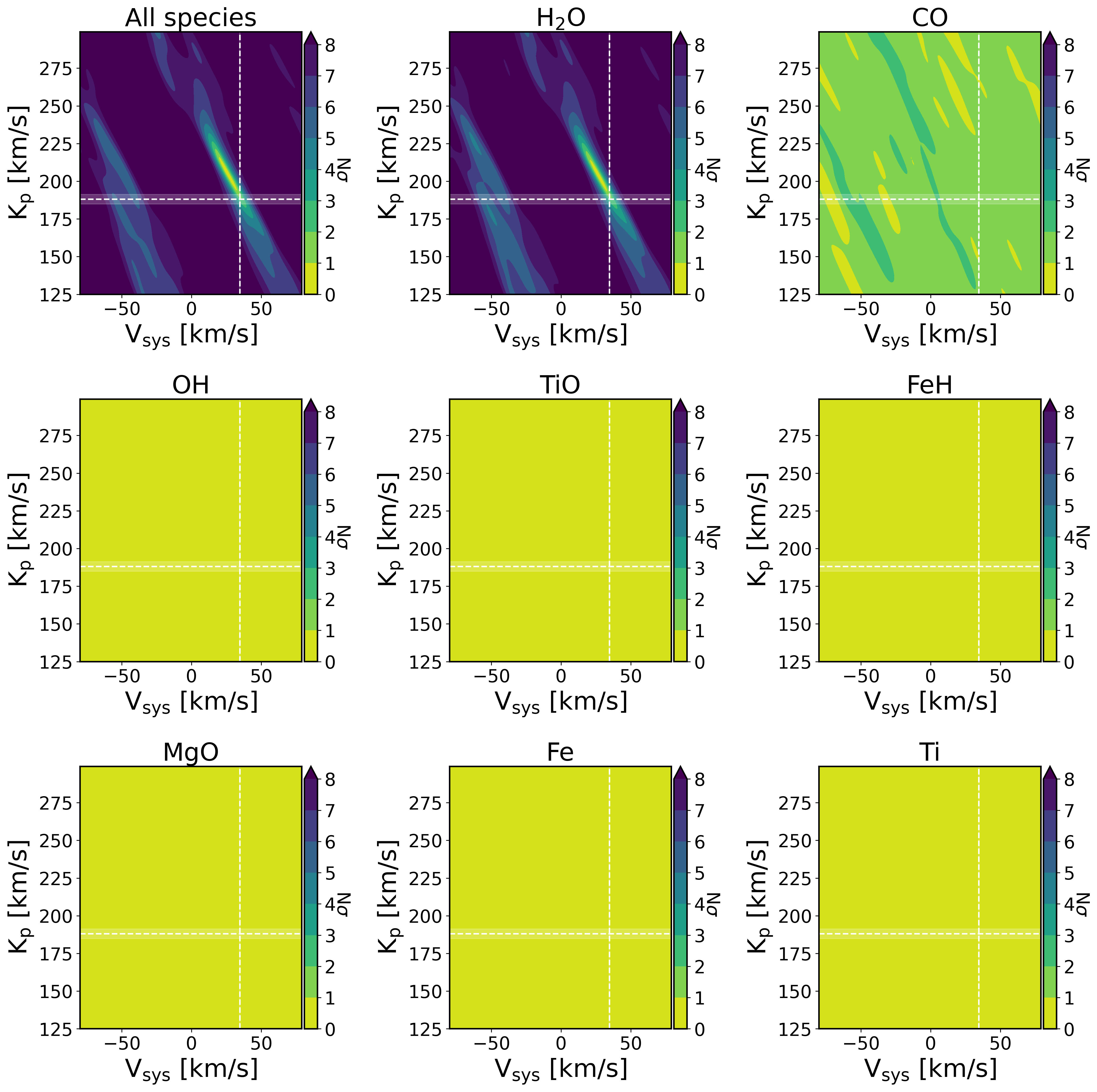}
\caption{Log-likelihood derived confidence interval maps for individual species corresponding to their best fit abundances from the equilibrium chemistry retrieval described in Section \ref{sec:retrieval_analysis}. We find that except \HtwoO{}, none of the other species have any significant contribution to planetary signal detected via cross-correlation.}
\label{fig:kpvsys_all_species}
\end{figure*}

We follow the same steps as \cite{smith_roasting_2024} to assess the contribution from individual chemical species to the cross-correlation signal by producing the \KpVsys{} maps for each of them. In brief, we first compute a \KpVsys{} log-likelihood map corresponding to the best fit model with the abundances of all the species set to their best fit values. Next, for a given species, we compute a model where the abundance of that species is set to a very small value (log$_{10}$ VMR = -30), and compute the \KpVsys{} CCF map corresponding to this model. The log-likelihood map representing the contribution from that species by itself is then computed as the difference between the all species log-likelihood map and the log-likelihood map with that species removed. The log-likelihood confidence intervals for each species in the equilibrium chemistry retrieval are shown in Figure \ref{fig:kpvsys_all_species}. As expected, we find that the detected cross-correlation signal for the best fit retrieved model is primarily driven by \HtwoO{}, and none of the other species show any strong contributions by themselves assuming there are no significant velocity shifts between different species.   

% \section{Effecting of varying \Npca{} on the retrieved posteriors}
% \label{app:npca_vs_retrieved_posteriors}
% \response{We tested the effect of varying \Npca{} on the retrieved posteriors from equilibrium chemistry retrieval. The retrieved posteriors for \KpVsys{} velocities, metallicity, and C/O are shown in Figure \ref{fig:npca_vs_posteriors}.

% \begin{figure*}
% \centering
% \includegraphics[width=1\textwidth]{figures/N_PCA_vs_posteriors.png}
% \caption{Comparison of the retrieved posteriors from different number of \Npca{} for equilibrium chemistry. The vertical dashed lines in the top two panels mark the predicted \Kp{} and \Vsys{} values from the literature.}
% \label{fig:npca_vs_posteriors}
% \end{figure*}

%%%%%%%%%%%%%%%%%%%%%%%%%%%%%%%%%%%%%%%%%%%%%%%%%%

% Don't change these lines
\bsp	% typesetting comment
\label{lastpage}
\end{document}